\documentclass[fleqn,usenatbib]{mnras}

\usepackage{newtxtext,newtxmath}

\usepackage[T1]{fontenc}
\usepackage{ae,aecompl}

\usepackage{graphicx}	
\usepackage{amsmath}	
\usepackage{amssymb}	
\usepackage{subfigure}
\usepackage{color}

\title[BH magnetosphere with small scale flux tubes]{Black hole magnetosphere with small scale flux tubes}

\author[Y. Yuan et al.]{
Yajie Yuan,$^{1}$\thanks{Lyman Spitzer, Jr. Postdoctoral Fellow.}\thanks{E-mail: yajiey@astro.princeton.edu (YY)}
Roger D. Blandford,$^{2}$
Dan R. Wilkins$^{2}$\thanks{NASA Einstein Fellow.}
\\
$^{1}$Department of Astrophysical Sciences, Princeton University, Princeton, NJ 08544, USA\\
$^{2}$Kavli Institute for Particle Astrophysics and Cosmology (KIPAC), Stanford University, Stanford, CA 94305, USA
}

\date{Accepted XXX. Received YYY; in original form ZZZ}

\pubyear{2018}

\begin{document}
\label{firstpage}
\pagerange{\pageref{firstpage}--\pageref{lastpage}}
\maketitle

\begin{abstract}
There is observational evidence that the X-ray continuum source that creates the broad fluorescent emission lines in some Seyfert Galaxies may be compact and located at a few gravitational radii above the black hole. We consider the possibility that this compact source may be powered by small scale flux tubes near the black hole that are attached to the orbiting accretion disk. As a first step, this paper investigates the salient features of black hole magnetospheres that contain small scale, disk-hole linking ``closed'' flux tubes, using the force-free approximation in an axisymmetric setting. We find that the extent of the closed zone is a result of the balance between the black hole spin induced twist in the closed zone and the confinement pressure of the external (open) field of the disk. The maximal extent of the closed zone, for a typical external confinement, is usually a few gravitational radii. The pressure competition between the closed zone and the external confinement could in principle lead to interesting dynamics and dissipation relevant for the compact X-ray corona.
\end{abstract}




\section{Introduction}

Many Active Galactic Nuclei (AGN) and X-ray binaries with stellar mass black holes show prominent, highly variable X-ray emission \citep[e.g.,][]{Elvis1994ApJS...95....1E,Grupe2010ApJS..187...64G,RemillardMcClintock2006ARA&A..44...49R}. 
This is usually believed to come from a hot ``corona'' around the accretion disk, but the mechanism for the formation and heating of the corona remains a mystery. 
Recently, significant progress has been made in understanding the geometric properties of coronae in Seyfert Galaxies, especially the ``bare'' ones that have little absorption along the line of sight to the nucleus. 
These Seyferts are typically spiral galaxies, with a central supermassive black hole of mass $10^6-10^7\;M_{\odot}$. The total luminosity  of the nucleus is in the range $\sim 0.03-1\;L_{\rm Edd}$, and the X-ray luminosity can be comparable to the optical/UV luminosity. They are radio quiet without strong jets, and as the luminosity is in the intermediate range, it is inferred that the accreted material forms a thin disk. 
The X-ray emission has two components: one is the hard X-ray power law continuum coming directly from the corona, due to energetic electrons in the corona upscattering disk thermal photons (mostly optical/UV) into the X-ray band; the other is the reflected (reprocessed) component, when the optically thick disk is irradiated by the X-ray continuum and produces fluorescent emission, as well as backscattering and secondary radiation \citep{Ross2005MNRAS.358..211R}. 

Several pieces of evidence have suggested that the X-ray corona is compact and located relatively close to the black hole, seemingly consistent with the so-called ``lamppost'' geometry often invoked in simplified modeling. 
Firstly, X-ray reverberation has been measured for dozens of sources, where it is possible to detect the time lags of the components in reflected spectrum as they respond to the change of the direct continuum emission \citep[e.g.,][]{Uttley2014A&ARv..22...72U,Kara2016MNRAS.462..511K}. 
Interpreting the time lag as due to the difference in light travel paths, it turns out that the irradiating source---the corona---is located within $\sim10$ gravitational radii ($r_g\equiv GM/c^2$) above the black hole. 
In addition, it has been shown that modeling the emissivity profile (rest frame emissivity as a function of the disk radius) of the prominent lines can give some constraints on the radial and vertical extent of the corona \citep{Wilkins2011MNRAS.414.1269W}. 
For example, \citet{Wilkins2015MNRAS.449..129W} found that during the high flux epoch of the narrow line Seyfert 1 galaxy Markarian 335, the corona has expanded, covering the inner region of the disk out to a radius of $26_{-7}^{+10}r_g$; while in the intermediate and low flux epochs, the corona contracts to within $\sim12r_g$ and $\sim5r_g$, respectively. Most interestingly, an X-ray flare was caught during the low flux epoch. The emissivity profile modeling suggests that, during the flare, the corona became collimated and extended vertically as if a jet launching event was aborted. After the flare, the corona had reconfigured into a much more compact form, within just $2-3\;r_g$ of the black hole. Such a dynamic sequence is quite suggestive as to what is powering the X-ray source. 
Last but not least, microlensing measurements of a few quasars also favor a compact X-ray corona located at a few gravitational radii above the black hole \citep[e.g.,][]{Morgan2008ApJ...689..755M,Chartas2009ApJ...693..174C,Mosquera2013ApJ...769...53M,ReisMiller2013ApJ...769L...7R}.

The formation and dynamics of the compact X-ray corona may be closely related to the physics underlying the dichotomy of radio loud and radio quiet AGN. 
GRMHD simulations have shown that in order to form powerful jets, the following conditions need to be satisfied \citep{McKinney2009MNRAS.394L.126M}: (1) a black hole with high spin; (2) a large-scale, coherent, dipole-type magnetic field threading the black hole and the disk; (3) good collimation, e.g. a thick accretion disk. 
Since the spin of black holes in some of the Seyferts have been measured to reach very high values (though there are caveats regarding the assumptions made in modeling), spin may be necessary but is clearly insufficient for a strong jet. Regarding the collimation, even with a thin disk it is still possible to have collimation from the disk wind or a strong field threading the disk \citep[e.g.][show that a thin disk, when threaded by large scale ordered field, can also become magnetically arrested and form a weak jet]{Avara2016MNRAS.462..636A}. So in the following we will focus on the magnetic field configuration near the black hole.

In particular, we consider the possibility that the field on the disk may be quite inhomogeneous, with small scale flux tubes emerging from the disk as well as patches of open flux bundles of different polarities and different coherent length scales \citep[e.g.][]{Blandford2002luml.conf..381B}.
Depending on the orientation of the field lines, they may or may not have much mass loading from the disk (\S\ref{sec:mass loading}).
Due to the continuous shear, the closed flux tubes connecting different radii of the disk may eventually be opened up, leading to current sheet formation and some dissipation due to reconnection.
The region near the black hole is of particular interest as the flux tubes would be able to wrap around the axis---the way they tangle up and untangle may involve continuous reconnection that can dissipate a significant fraction of the Poynting flux flowing along the flux tubes. 
Also, in this region the flux tubes can be connected to the black hole and extract the rotational energy of the hole, which may be able to power the X-ray source. 

In what follows, we first write down a general formalism to describe the mass loading on the field lines from a thin accretion disk due to centrifugal and gravitational effects (\S\ref{sec:mass loading}). We then use simple force-free models to illustrate some salient features of black hole magnetospheres with small scale flux tubes (\S\ref{sec:solutions} and \S\ref{sec:ELflux and emissivity}). This class of toy models have flux bundles linking the black hole and the disk, resembling those studied by \citet{Uzdensky2005ApJ...620..889U,Parfrey2015MNRAS.446L..61P}. The goal is to demonstrate the equilibrium condition for such a flux bundle based on its interaction with surrounding fields, in an axisymmetric setting. The rotating black hole can be regarded as a battery and the force-free magnetosphere a circuit driven by the EMF. While large scale, ordered fields are more likely to be connected to the formation of jets and outflows whereby the load of the circuit is located at very large distances, the small scale flux tubes linking the hole and the disk could have load much closer in, e.g. on the disk or at a dissipation site above the disk where reconnection takes place. We discuss the implications of these solutions in \S\ref{sec:discussion} and conclusions in \S\ref{sec:conclusion}.  

\section{Mass loading of the field lines}\label{sec:mass loading}
In a highly magnetized plasma, the motion of charged particles is confined to the field lines. A good analogy here is to imagine the magnetic field lines as wires and the particles as beads moving freely along the wires. \citet{Blandford1982MNRAS.199..883B} showed that for field lines emanating from a thin disk around a massive object with Newtonian gravitational potential, if the poloidal component of the magnetic field makes an angle $<60^{\circ}$ with respect to the outward radius of the disk surface, the matter on the disk surface will be flung outwards. In the corotating frame, this region has a centrifugal force larger than the gravitational force. Meanwhile, another unstable region exists: when the poloidal component of the field makes an angle $<60^{\circ}$ with respect to the inward radius of the disk surface, matter will slide inwards. In this region, the gravitational force wins over the centrifugal force. In general, field lines lying in these two regions will naturally have mass loading from the disk, even if the gas is cold. Note that the conclusion does not depend on the toroidal component of the magnetic field.

The same analysis can be generalized to the spacetime around a Kerr black hole. Imagine a wire/field line anchored on a disk, the foot point orbiting with the same angular velocity $\Omega$ as the disk material there \footnote{We shall mostly use a Keplerian $\Omega$, appropriate to a thin disk. Thick disks have smaller $\Omega$ but the principles are unchanged.}. Now if we make the following transformation of variables to switch to the corotating point of view (in what follows, variables with bars are represented using corotating coordinates, while those without bars are in Boyer-Lindquist coordinates)
\begin{equation}\label{eq:transformation}
\bar{t}=t,\ \bar{\phi}=\phi-\Omega t,
\end{equation}
the wire/magnetic field line will appear stationary. Parametrize the curve as $\bar{\ell}^{\mu}(s)$, where $s$ measures the distance along the curve, and $\bar{\ell}^{0}(s)=0$. The 4-velocity of a particle moving along the wire can then be written as
\begin{equation}
\bar{u}^0=\frac{dt}{d\tau},\ 
\bar{u}^{j}=\frac{d\bar{\ell}^{j}}{ds}\frac{ds}{d\tau}\equiv \bar{b}^{j}\frac{ds}{d\tau}.
\end{equation}
So the Lagrangian of the particle can be obtained as
\begin{equation}
\bar{L}=\frac{1}{2}\bar{g}_{\mu \nu } \frac{d \bar{x}^{\mu }}{d\tau} \frac{d \bar{x}^{\nu }}{d\tau }=\frac{1}{2}\bar{g}_{00}\left(\frac{dt}{d\tau}\right)^2+\bar{g}_{0j}\bar{b}^j\frac{dt}{d\tau}\frac{ds}{d\tau}+\frac{1}{2}\bar{b}^2\left(\frac{ds}{d\tau}\right)^2.
\end{equation}
We note that under the transformation of Equation (\ref{eq:transformation}), the metric tensor becomes
\begin{equation}
\bar{g}_{\mu \nu }=\left(
\begin{array}{cccc}
 g_{tt}+2 \Omega  g_{t\phi }+\Omega ^2 g_{\phi \phi } & 0 & 0 & g_{0 \phi }+\Omega  g_{\phi \phi } \\
 0 & g_{rr} & 0 & 0 \\
 0 & 0 & g_{\theta \theta } & 0 \\
 g_{0 \phi }+\Omega  g_{\phi \phi } & 0 & 0 & g_{\phi \phi } \\
\end{array}
\right).
\end{equation}
$\bar{g}_{\mu\nu}$ becomes singular on the light surface where $\bar{g}_{tt}=g_{tt}+2 \Omega  g_{t\phi }+\Omega ^2 g_{\phi \phi }=0$. Nevertheless this is only a problem with the pathological coordinate choice; the particle motion remains fully physical. Back to Boyer-Lindquist coordinates, and noticing $\bar{b}^0=b^0=0$, $\bar{b}^{\phi }=b^{\phi }-b^0 \Omega=b^{\phi}$, the Lagrangian becomes
\begin{align}
L=\bar{L}&=\frac{1}{2}(g_{tt}+2 \Omega  g_{t\phi }+\Omega ^2 g_{\phi \phi })\left(\frac{dt}{d\tau}\right)^2\nonumber\\
&+(g_{0 \phi }+\Omega  g_{\phi \phi })b^{\phi}\frac{dt}{d\tau}\frac{ds}{d\tau}+\frac{1}{2}b^2\left(\frac{ds}{d\tau}\right)^2.
\end{align}
The generalized momenta are
\begin{align}
\pi_t&=\frac{\partial L}{\partial (dt/d\tau)}=\frac{\partial \bar{L}}{\partial (dt/d\tau)}=\bar{u}_0\nonumber\\
&=(g_{tt}+2 \Omega  g_{t\phi }+\Omega ^2 g_{\phi \phi })\left(\frac{dt}{d\tau}\right)+(g_{0 \phi }+\Omega  g_{\phi \phi })b^{\phi}\frac{ds}{d\tau},\\
\pi_s&=(g_{0 \phi }+\Omega  g_{\phi \phi })b^{\phi}\frac{dt}{d\tau}+b^2\frac{ds}{d\tau}.
\end{align}
And the Hamiltonian can be written as
\begin{align}
H&=\pi_t\frac{dt}{d\tau}+\pi_s\frac{ds}{d\tau}-L
=-\frac{\pi _t^2 b^2-2D \pi _t \pi _s+K \pi _s^2}{2 \left(D^2-b^2 K\right)}\nonumber\\
&=-\frac{K\left(\pi_s-\frac{D}{K}\pi_t\right)^2}{2(D^2-Kb^2)}+\frac{\pi_t^2}{2K},
\end{align}
where $K=g_{tt}+2 \Omega  g_{t\phi }+\Omega ^2 g_{\phi \phi }$ and $D=(g_{0\phi}+g_{\phi\phi}\Omega)b^{\phi}$.
Since $H$ does not depend explicitly on $t$, $\pi_t$ is conserved. We can regard $\pi_t=\bar{u}_0=u_0+\Omega u_{\phi}$ as the ``energy'' in the corotating frame \citep[see also][for a discussion of conserved quantities for a particle moving on the field sheet]{Gralla2014MNRAS.445.2500G}. Meanwhile, $H$ itself is conserved, 
\begin{equation}
H=-\frac{K\left(\pi_s-\frac{D}{K}\pi_t\right)^2}{2(D^2-Kb^2)}+\frac{\pi_t^2}{2K}=-\frac{1}{2}.
\end{equation}
Now in the region where a physical corotating frame exists, $K<0$, $D^2-Kb^2>0$, so we can regard the first term in the above expression as the effective kinetic energy and the second term as the effective potential energy:
\begin{equation}
V_{\rm eff}=\frac{\pi_t^2}{2K}.
\end{equation}
The light surfaces where $K=0$ mark a transition in the characteristics of the motion of the particle: outside the outer light surface, particles moving both ways along the wire/magnetic field appear to be moving outward and cannot return to the inner region; similarly, inside the inner light surface, particles moving both ways along the wire/magnetic field appear to be moving inward and cannot escape to the outer region.

\begin{figure}
	\centering
	\includegraphics[width=0.925\columnwidth]{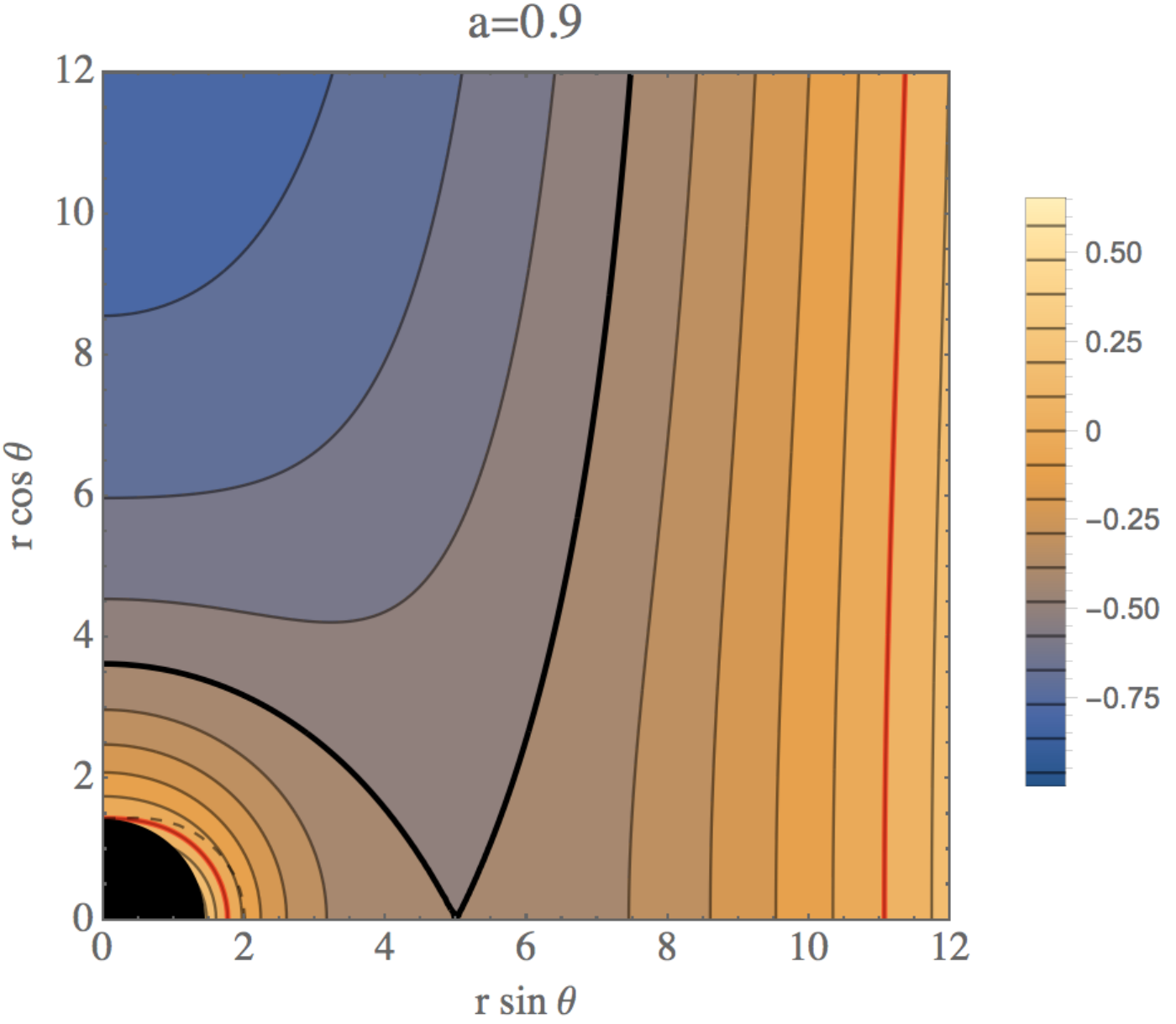}
    \caption{Contour plot of the function $K=g_{tt}+2 \Omega  g_{t\phi }+\Omega ^2 g_{\phi \phi }$, for the case where the black hole spin is $a=0.9$ and the field line is rotating with the prograde Keplerian angular velocity at $r_d=5$ (the length is measured in $M$, or equivalently, in $r_g\equiv GM/c^2$; same below). The black hole is shown by the black sphere and the boundary of the ergosphere is shown by the black dashed line. The thick red lines indicate the location where $K=0$. The thick black contour separates the stable region and the unstable regions.}
    \label{fig:K-a0.9r5}
\end{figure}

\begin{figure}
	\centering
	\includegraphics[width=0.9\columnwidth]{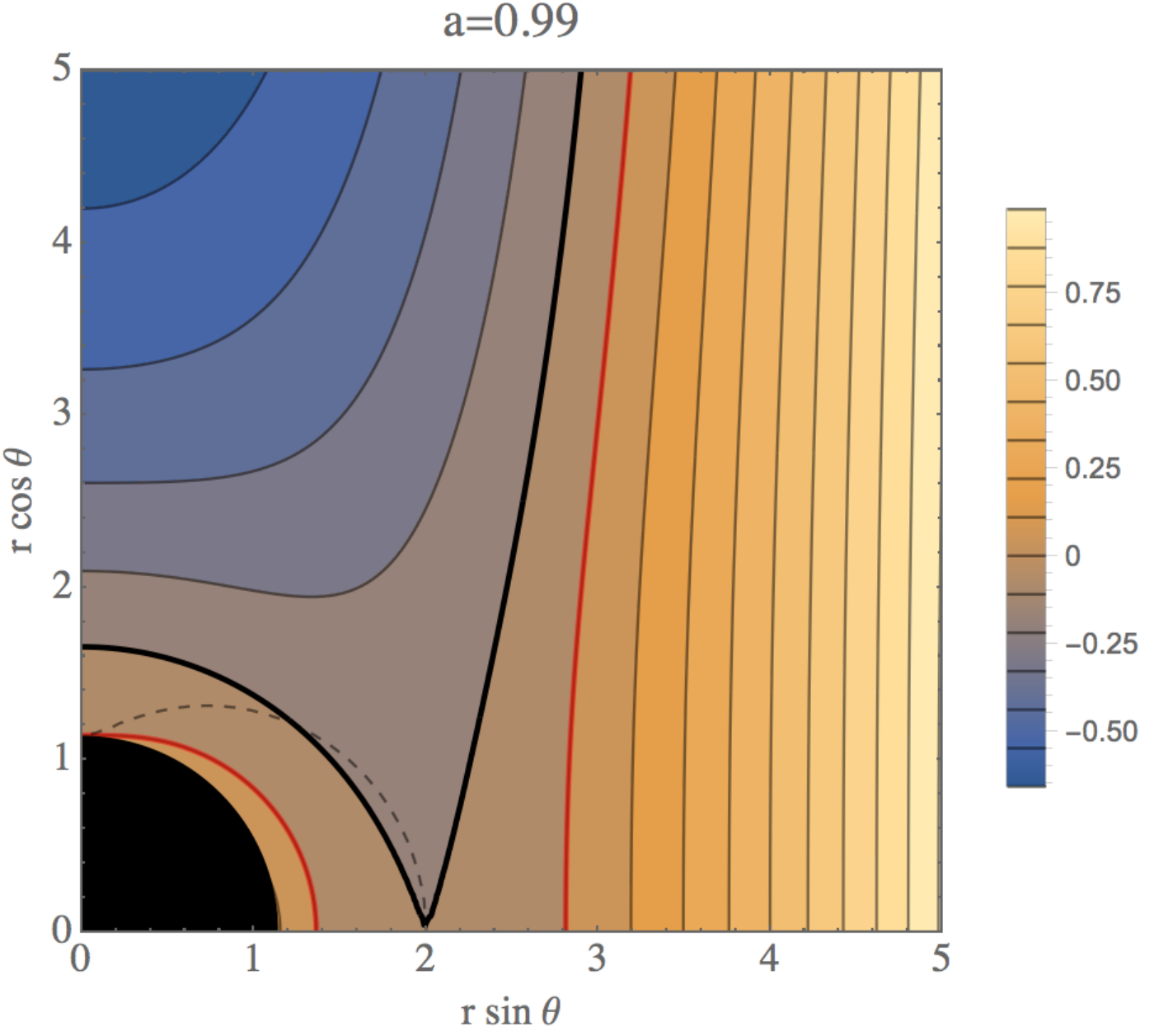}
    \caption{Similar to Figure \ref{fig:K-a0.9r5} except that the black hole spin is $a=0.99$ and the field line is rotating with the prograde Keplerian angular velocity at $r_d=2$.}
    \label{fig:K-a0.99r2}
\end{figure}

\begin{figure}
	\centering
	\includegraphics[width=0.9\columnwidth]{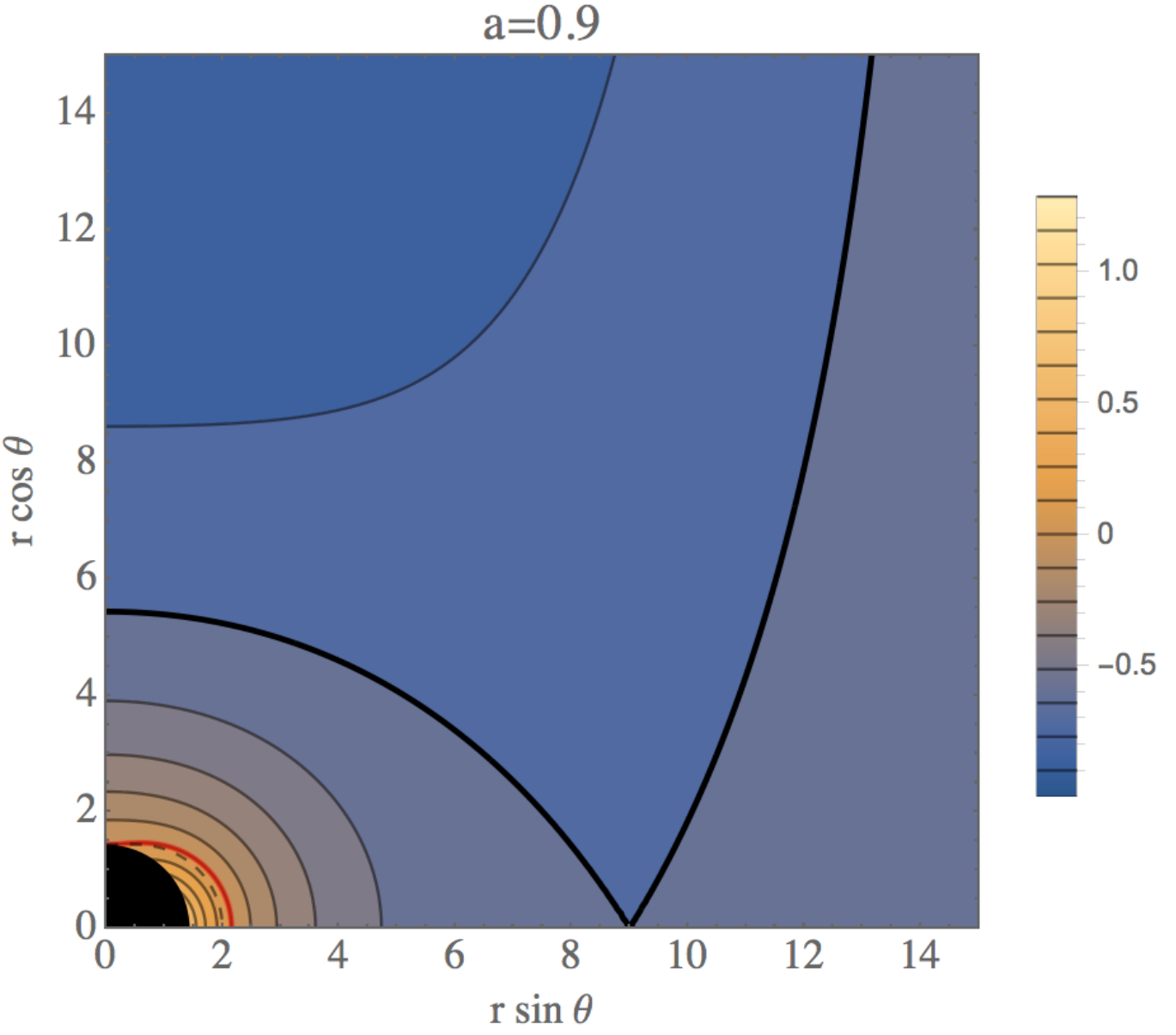}
    \caption{Similar to Figure \ref{fig:K-a0.9r5} except that the black hole spin is $a=0.99$ and the field line is rotating with the retrograde Keplerian angular velocity at $r_d=9$.}
    \label{fig:K-a0.9r9retro}
\end{figure}

\begin{figure}
	\centering
    \includegraphics[width=0.9\columnwidth]{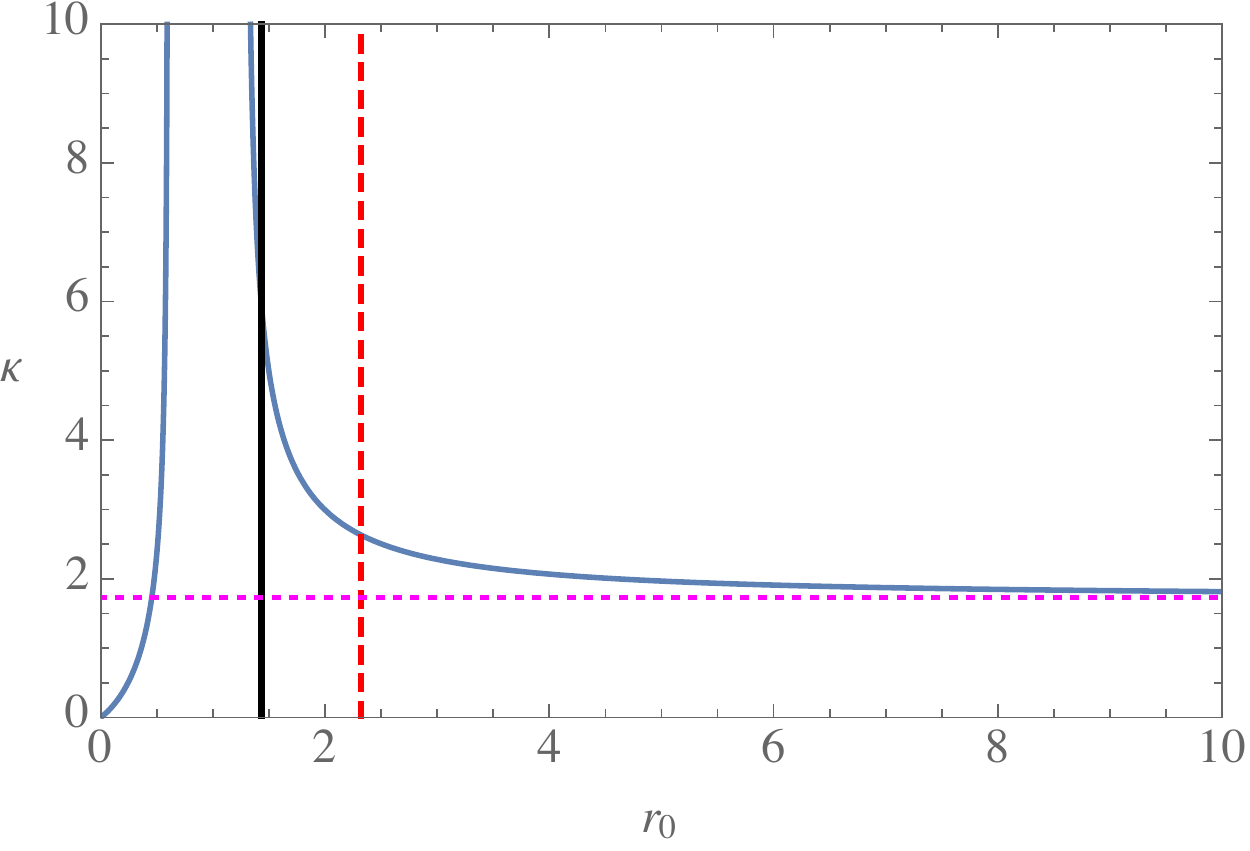}
    \caption{The blue line shows the critical quantity $\kappa\equiv (r_db^{\theta}/b^r)_{cr}$ at the field line foot point as a function of $r_d$, for prograde orbits around a Kerr black hole with $a=0.9$. For field lines with $r_db^{\theta}/b^r<\kappa$, the bead at the foot point is unstable. The thick black line is the radius of the event horizon and the red dashed line is the radius of ISCO. The magenta dotted line is the asymptotic non-relativistic value $\sqrt{3}$.}
    \label{fig:angle_a0.9}
\end{figure}

\begin{figure}
	\centering
    \includegraphics[width=0.9\columnwidth]{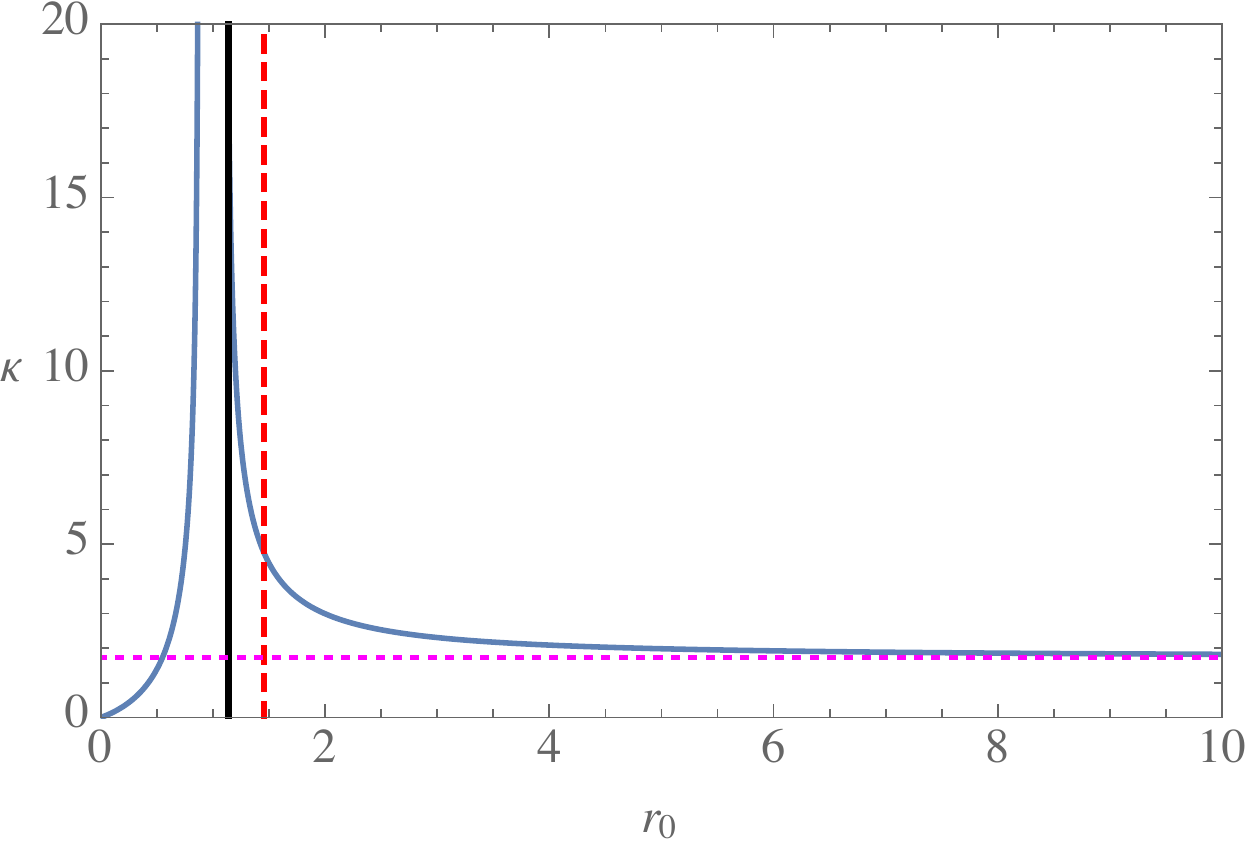}
    \caption{Similar to Figure \ref{fig:angle_a0.9} but with $a=0.99$. $\kappa$ remains finite at the event horizon. In this particular case, $\kappa=17.87$ at the event horizon.}
    \label{fig:angle_a0.99}
\end{figure}

In Figures \ref{fig:K-a0.9r5}, \ref{fig:K-a0.99r2} and \ref{fig:K-a0.9r9retro} we show examples of contour plots of the function $K=\pi_t^2/(2V_{\rm eff})=g_{tt}+2 \Omega  g_{t\phi }+\Omega ^2 g_{\phi \phi }$. It can be seen that if the wire/field line lies in the yellow regions either to the left of the leftmost thick separating contour or to the right of the rightmost thick separating contour, the effective potential decreases along the field line, so the matter will have a tendency to move away from the disk. For the field lines lying in the blue region in between the thick separating contours, the effective potential increases along the field line and the matter tends to stay at the foot point of the field line. If we expand the potential near the foot point ($r=r_d$, $\theta=\pi/2$), noticing that the angular velocity is $\Omega=(a+r_d^{3/2}/\sqrt{M})^{-1}$, we get
\begin{align}
V_{\text{eff}}&=Const.+\frac{M \pi _t^2 \left(a \sqrt{M}+r_d^{3/2}\right)^2}{2 r_d^3 \left(2 a \sqrt{M}+\sqrt{r_d} \left(r_d-3 M\right)\right)^2}\times\nonumber\\
&\left(\left(b^{\theta }\right)^2 \left(3 a^2-4 a \sqrt{M} \sqrt{r_d}+r_d^2\right)-3 \left(b^r\right)^2\right)s^2+O\left(s^3\right).
\end{align}
So the bead is unstable if
\begin{equation}
\left(b^{\theta }\right)^2 \left(3 a^2-4 a \sqrt{M} \sqrt{r_d}+r_d^2\right)-3 (b^r)^2<0. 
\end{equation}
When $a=0$ this condition is $(r_d b^{\theta })^2<3 (b^r)^2$, at large $r_d$ this coincides with the Newtonian result at the foot point. For larger $a$, the critical angle changes as we get close to the event horizon. Figures \ref{fig:angle_a0.9} and \ref{fig:angle_a0.99} show a few examples of the critical quantity $\kappa\equiv (r_db^{\theta}/b^r)_{cr}$ for different black hole spins. It can be seen that $\kappa$ increases (but remains finite) when the foot point gets close to the event horizon, meaning that the range of the unstable region increases, especially when $a$ approaches unity, in agreement with \citet{Lyutikov2009MNRAS.396.1545L}.

The applicability of the effective potential is not limited to field lines that emanate from the disk; it can be used anywhere the angular velocity of the field line is known, e.g., field lines threading the black hole event horizon. In a more self-consistent MHD formalism, \citet{Takahashi1990} found that a 1D cold flow on a flux tube connecting the event horizon and infinity possesses a stagnation point where matter has to be injected; if the injected material has zero poloidal velocity to start with, the injection point coincides exactly with the maximum of the effective potential $V_{\rm eff}$ along the field line. After all, the simple model of beads on a wire describes correctly the behavior of particles along the field line when the shape of the field line is fixed.    

\section{A few examples of force-free solutions}\label{sec:solutions}
In the following we look at a particular class of magnetospheric configurations where all the field lines that go through the black hole event horizon connect to the disk. For simplicity we consider the force-free limit, which should be a good approximation if the mass loading on the field lines is small such that the magnetization parameter $\sigma\equiv B^2/(4\pi n m c^2)$ is sufficiently large. Some of these have been investigated by \citet{Uzdensky2005ApJ...620..889U} \citep[see also discussion in][]{2010mfca.book.....B}. Here, we are going to look into the detailed properties, including mass loading on the field lines, emissivity profile of the disk, as well as their dependence on the disk magnetic flux distribution and black hole spin (especially at high spins). We will then discuss possible dissipation of the magnetic energy around the separatrix point in 3D settings.

\subsection{Basic properties}

\begin{figure}
	\centering
    \subfigure[]
    {
    	\includegraphics[width=\columnwidth]{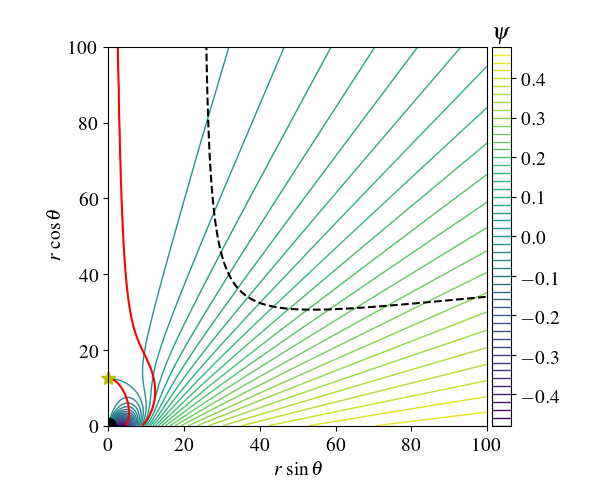}
    }
    \subfigure[]
    {
    	\includegraphics[width=\columnwidth]{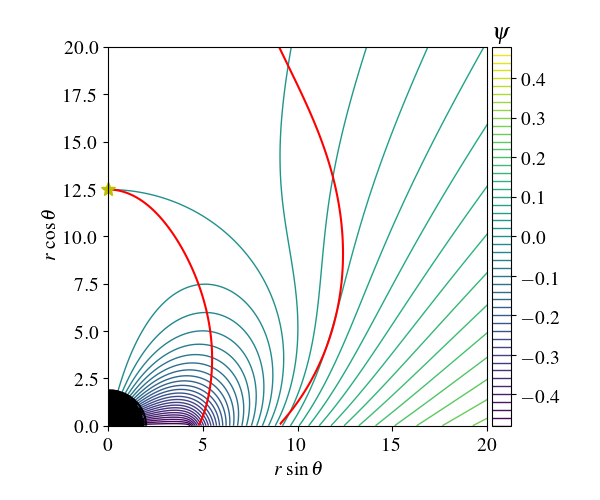}
    }
    \caption{Force-free solution for the case of a Kerr black hole with $a=0.5$ and a thin disk orbiting at Keplerian angular velocity up to the ISCO. The upper panel shows the large scale field structure and the lower panel is a zoom-on view near the black hole. The field is monopole-like at large distances. The color contours show the flux function $\psi$. The black sphere corresponds to the black hole event horizon. The black solid line is the boundary of the ergosphere (too close to the event horizon to be distinguishable on the plot) and the black dashed lines are the light surfaces (the inner light surface lies inside the ergosphere). The two red lines mark the extrema of $V_{\rm eff}$ along field lines. The yellow star marks the location of the null point where $B=0$.}
    \label{fig:rinverse-a0.5}
\end{figure}

\begin{figure}
	\centering   
    \includegraphics[width=0.8\columnwidth]{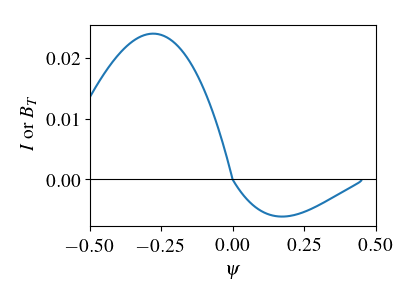}
    \caption{The poloidal current $I$ or equivalently $B_T\equiv$ the toroidal magnetic field as a function of $\psi$ for the solution shown in Figure \ref{fig:rinverse-a0.5}. We assume the poloidal magnetic field is pointing away from the disk in both the closed field line region and open field line region. The enclosed poloidal current $I$ is measured from the pole. The separatrix (last closed field line) corresponds to $\psi=0$; $\psi$ decreases toward $-0.5$ as the field line foot point gets close to the ISCO, and $\psi$ increases toward $0.5$ as the foot point approaches very large radius on the disk.}
    \label{fig:rinverse-a0.5-I}
\end{figure}

As our first example, we consider a thin disk orbiting at prograde Keplerian angular velocity $\omega=(a+r^{3/2}/\sqrt{M})^{-1}$, up to the innermost stable circular orbit (ISCO). The field lines are anchored on the disk and rotate with the same angular velocity as their foot point. We assume that field lines emanating from $r<r_0$ on the disk connect to the black hole (closed zone), while those emerging from $r>r_0$ open to infinity (open zone). Here $r_0$ is a free parameter in our initial condition; our numerical solver will determine whether such an initial condition leads to a final steady state. The field line that has its foot point exactly at $r_0$ is special: it connects to the polar axis at a point where $B=0$---the null point, so has the same flux function value $\psi$ as the axis. It is a separatrix: separating the closed zone and the open zone. As boundary conditions, we set $\psi$ to be zero on the pole ($\theta=0$); on the disk it is fixed to be $\psi \left(r,\theta =\pi /2\right)=f(r)=r_{\text{ISCO}}/r_0-r_{\text{ISCO}}/r$ for $r>r_{\rm ISCO}$ so the separatrix has $\psi=0$ as well. The boundary condition on the equatorial plane inside the ISCO (``plunging region'', $\theta=\pi/2$, $r<r_{\rm ISCO}$) is set to be a constant: $\psi=f(r_{\rm ISCO})$, namely, the magnetic field lies along the equatorial plane in the plunging region. Our setup is essentially the same as \citet{Uzdensky2005ApJ...620..889U}, except that instead of treating open field lines as potential field with zero angular velocity in \citet{Uzdensky2005ApJ...620..889U}, we allow all field lines to be orbiting with Keplerian angular velocity of their foot points, namely, the whole region outside the disk is force-free. The solution is obtained using a relaxation procedure as described in Appendix \ref{sec:method}. 

Figure \ref{fig:rinverse-a0.5} shows the axisymmetric, steady state solution for the case $a=0.5$ and the last closed field line located at $r_0=2r_{\rm ISCO}$. The poloidal current $I$ within the flux surface, or, equivalently, the quantity $B_T\equiv$ the toroidal magnetic field, as a function of $\psi$, determined self-consistently from the smoothness condition at the light surfaces, is shown in Figure \ref{fig:rinverse-a0.5-I}. On the separatrix surface (last closed flux surface), the toroidal field is zero, and there is no surface current separating the closed zone and the open field zone.

A few interesting features are worth noting here. Firstly, the steady state force-free equation does not constrain the sign of the current/toroidal field, so some care is needed to choose the physical solution. Suppose the poloidal magnetic field is directed away from the disk for both the open field lines and the closed field lines. In the open field line region, the magnetic field lines follow the disk rotation, and are swept back around and beyond the light surface. So the toroidal magnetic field is negative as required by causality. Here the poloidal current density flows into the disk around the pole but a volumetric return current (leaving the disk) exists at larger $\theta$. In the closed field line region, since the black hole event horizon angular velocity $\Omega_H$ is larger than the disk angular velocity everywhere, and the magnetic field lines are dragged into rotation by the ergosphere, we should expect the field lines to be lagging with respect to the black hole rotation, namely, the toroidal field $B_T$ should be positive for the field lines that go into the black hole. This would mean that the field lines are leading at the disk surface, but it does not violate causality since there is Alfv\'{e}n wave communication between the disk and the ergosphere---the inner light surface lies well within the ergosphere. From figure \ref{fig:rinverse-a0.5-I} we can see that the current leaves the black hole near the pole; some return current enters the hole near the equator and there is a return current sheet on the equatorial plane between the event horizon and the ISCO.  

To look at possible mass loading on the field lines, we plot the locations of the extrema of the effective potential $V_{\rm eff}$ along field lines. These are shown by the red lines in Figures \ref{fig:rinverse-a0.5} and \ref{fig:a0.5compare}(a). Only field lines that lie completely inside the inner red line or outside the outer red line can naturally have mass loading on them from a cold disk. For these regions, a RMHD approach may be needed to determine the flow more self-consistently. On field lines where mass loading is unlikely, there can be other ways to populate the region with sufficient charge carriers, e.g. by a pair cascade near the black hole \citep{Blandford1977MNRAS.179..433B} or by external gamma ray pair production \citep[e.g.,][]{phinney_theory_1983}.

\subsection{Dependence on the disk magnetic flux distribution}

\begin{figure*}
	\centering
    \includegraphics[width=\textwidth]{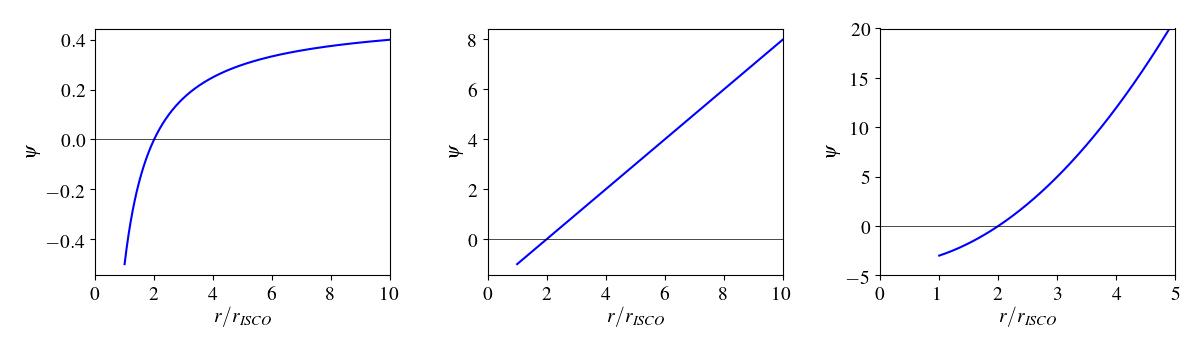}
    \caption{3 different types of flux distribution on the disk we considered. Left: monopole-like field at large distances; middle: asymptotically paraboloidal field; right: asymptotically vertical field.}
    \label{fig:diskflux}
\end{figure*}

\begin{figure*}
	\centering
    \includegraphics[width=\textwidth]{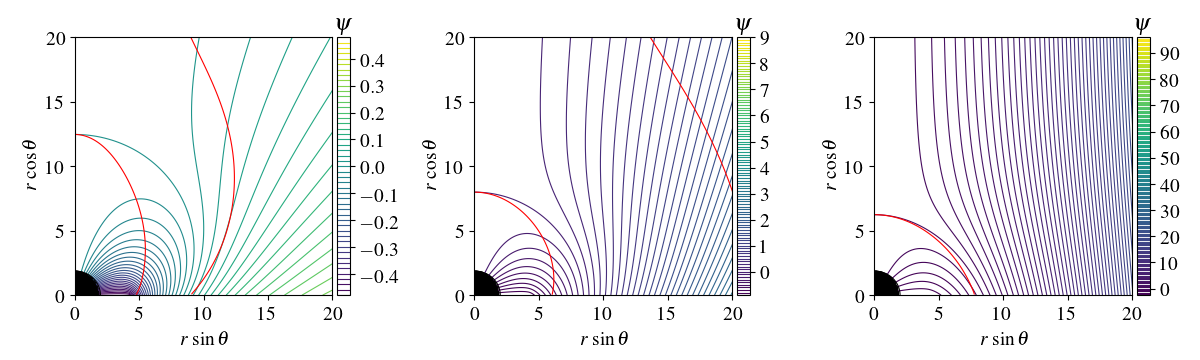}
    \caption{A comparison of magnetospheric solutions for 3 different types of flux distributions on the disk. Left: monopole-like field at large distances (a zoom-in view of Figure \ref{fig:rinverse-a0.5}); middle: asymptotically paraboloidal field; right: asymptotically vertical field. All three cases have $a=0.5$ and the last closed field line connects the disk at $r_0=2r_{\rm ISCO}$. The black sphere corresponds to the black hole event horizon. Both the ergosphere and the inner light surface are very close to the event horizon and not distinguishable on the plot. The red lines mark the extrema of $V_{\rm eff}$ along field lines. Note that the open field lines in the second and third cases are set to have zero angular velocity; the effective potential extrema on these open field lines are just for illustration if the field lines were orbiting with Keplerian angular velocity too.}
    \label{fig:a0.5compare}
\end{figure*}


\begin{figure*}
	\centering
	\includegraphics[width=\textwidth]{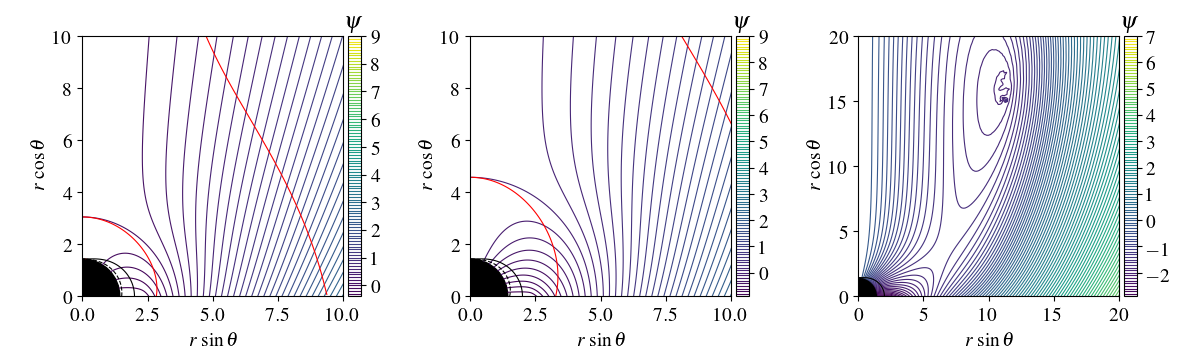}
    \caption{A comparison of different extent of the closed zone. All three cases have $a=0.9$ and the magnetic field is asymptotically paraboloidal at large distances. (a) The last closed line connects to the disk at $r_0=1.5r_{\rm ISCO}$. (b) $r_0=2r_{\rm ISCO}$. (c) Setting $r_0=4r_{\rm ISCO}$ we fail to reach a steady state solution: the figure shows the ejection of a plasmoid where the relaxation procedure breaks down.}
    \label{fig:a0.9compare}
\end{figure*}

\begin{figure*}
	\centering
    \includegraphics[width=\textwidth]{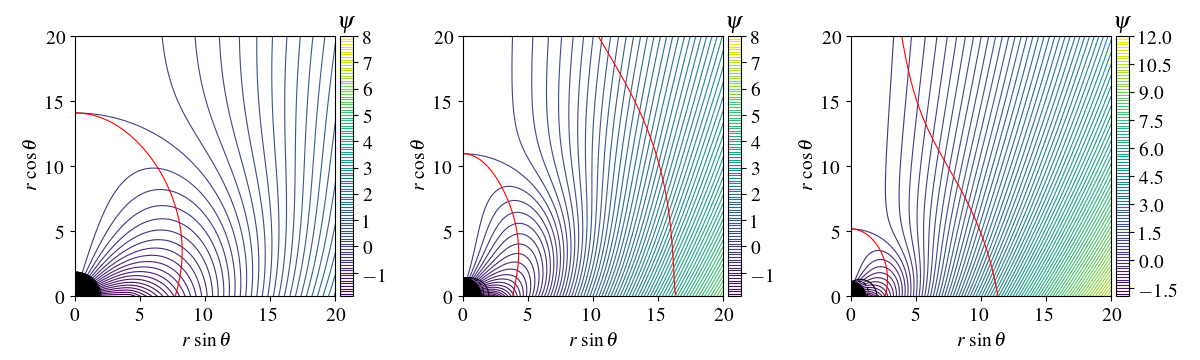}
    \caption{A comparison of different black hole spins. All three cases  have asymptotically paraboloidal magnetic field and the last closed line connects to the disk at $r_0=3r_{\rm ISCO}$. (a) $a=0.5$. (b) $a=0.9$. (c) $a=0.99$.}
    \label{fig:a-compare}
\end{figure*}


Firstly, we notice that the extent of the closed zone and the tendency of mass loading on these field lines clearly depend on the pressure from the open field zone. To illustrate this, we consider another two types of flux distribution on the disk. One has $\psi \left(r,\theta =\pi/2\right)=f(r)=\left(r-r_0\right)/r_{\text{ISCO}}$ for $r\geq r_{\text{ISCO}}$; this corresponds to asymptotically paraboloidal field. The other has $\psi \left(r,\theta =\pi/2\right)=f(r)=\left(r^2-r_0^2\right)/r_{\text{ISCO}}^2$ for $r\geq r_{\text{ISCO}}$, representing asymptotically uniform vertical field. All three cases of $\psi$ as a function of disk radius are shown in Figure \ref{fig:diskflux}. When getting the force-free solutions for the latter two cases, we find that the outer light surface typically lies at very large distances (or does not exist) and poses numerical difficulties. As an expedient measure, we set the angular velocity of the open field lines to be zero, namely, again treat them as potential field for these cases.

Figure \ref{fig:a0.5compare} shows the final steady state solution for three examples where the black hole spins are all $a=0.5$ and the last closed field line connects to the disk at $r_0=2r_{\rm ISCO}$, while the flux distributions on the disk correspond to the three different cases shown in Figure \ref{fig:diskflux}. It indeed shows that increasing the magnetic pressure from the open field lines can push down on the closed zone, leading to a smaller closed zone and more mass loading on the closed field lines.

\subsection{Dependence on black hole spin}
Another important point, as already discussed by \citet{Uzdensky2005ApJ...620..889U}, is that the black hole spin limits the radial extent of the force-free link on the disk surface. As the field lines rotate with the disk angular velocity, which is different from the angular velocity of the black hole event horizon, these field lines have to develop a toroidal component, in order for the field lines to slip across the event horizon. When the field lines from the polar region of the black hole connects to a very large radius on the disk, the black hole spin may induce a toroidal field with larger pressure than can be confined by the tension of the poloidal field, so a force-free equilibrium no longer exists.

Our results are consistent with this. Figure \ref{fig:a0.9compare} shows an example where the black hole spin is $a=0.9$ and the magnetic field is asymptotically paraboloidal. We try to initialize our relaxation procedure with the separatrix field line connecting to the disk at different $r_0$, and see if a steady state solution can be obtained. It turns out that for $r_0\le3r_{\rm ISCO}$, a final steady state exists, and the height of the null point increases with $r_0$, see Figure \ref{fig:a0.9compare} (a)(b) and Figure \ref{fig:a-compare} (b). But when $r_0=4r_{\rm ISCO}$, we can no longer get a steady solution: during the relaxation the disk-hole link tends to open up and eject plasmoids, causing the procedure to break down (Figure \ref{fig:a0.9compare}c). We find similar trend when $a=0.5$ and $a=0.99$ (with the same asymptotically paraboloidal open flux): the closed zone can extend to $r_0=3r_{\rm ISCO}$ but no longer exists when $r_0=4r_{\rm ISCO}$. Figure \ref{fig:a-compare} shows the three cases with different black hole spin but the same $r_0=3r_{\rm ISCO}$. In general, for higher black hole spins, the maximum extent of the closed zone gets smaller, both radially on the disk surface and vertically on the polar axis.

\begin{figure}
\centering
\includegraphics[width=\columnwidth]{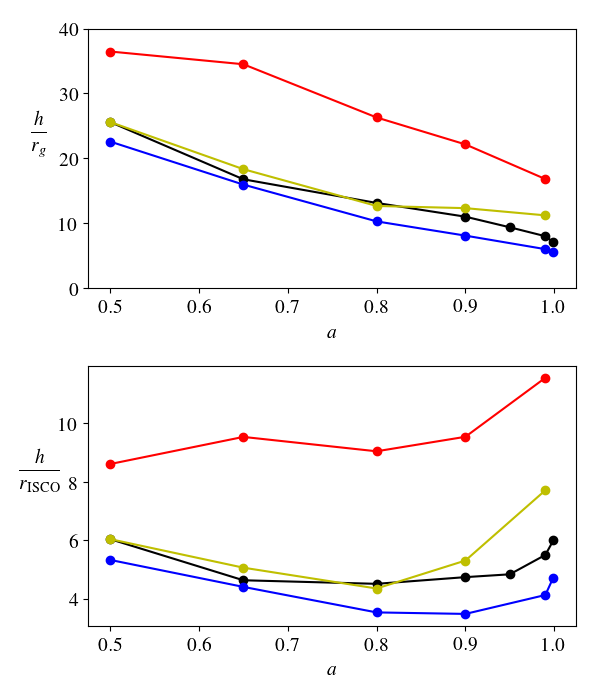}
\caption{Maximal height of the null point, for different spins and different flux distributions on the disk. The upper panel shows the height $h$ in terms of $r_g$, while the lower panel shows the ratio $h/r_{\rm ISCO}$. Black: the flux distribution is asymptotically paraboloidal and the disk inner boundary extends to $r_{\rm ISCO}$; yellow: the flux distribution is also asymptotically paraboloidal but the disk inner boundary are all fixed at $4.233r_g$ ($\ge r_{\rm ISCO}$ for $a\ge 0.5$); blue: the flux distribution is asymptotically monopolar and the disk inner boundary extends to $r_{\rm ISCO}$; red: the flux distribution is asymptotically vertical and the disk inner boundary extends to $r_{\rm ISCO}$.}\label{fig:height}
\end{figure}

There is a fundamental reason behind this. In order to establish a closed zone with field lines leading at the disk, the flow along the field lines has to be subsonic (sub-Alfv\'{e}nic in the force-free limit) such that characteristic information can travel along the field lines. When a section of a sheared field line gets farther and farther away from the disk, it becomes harder and harder for the information to propagate back and forth along the field line within the shearing time scale. When the communication cannot be maintained any more, the field line breaks and gets swept back at the disk.

To quantify the maximum extent of the closed zone, for a given flux distribution on the disk and a given black hole spin, we try to increase the separatrix location $r_0$ until a steady state solution no longer exists. The maximum height of the null point can then be recorded. Figure \ref{fig:height} shows the results. We can see that for a given flux distribution, there is a robust trend that the maximum height of the null point decreases with the black hole spin. Increasing the pressure from the open field lines can allow the closed zone to extend to larger height, as demonstrated by a comparison among different flux distributions. These results suggest that the extent of the closed zone is a consequence of the combined effects of the black hole spin that twist up the fields and the tension/pressure from both the poloidal field inside the closed zone and the open field lines.

\section{Feedback on disk motion and emissivity profile}\label{sec:ELflux and emissivity}
The field lines connecting the black hole and the disk allow angular momentum and energy transfer between the hole and the disk. In the following we look at its influence on the disk motion and emissivity profile.

\subsection{Energy and angular momentum flux from the black hole to the disk}
Using Killing vectors $\chi=\partial/\partial t$ and $\eta=\partial/\partial \phi$, we can get the energy flux $\mathcal{E}^{\mu }=-\chi _{\nu } T^{\mu \nu }=-T_0^{\mu }$ and the angular momentum flux $\mathcal{L}^{\mu }=\eta _{\nu } T^{\mu \nu }=T_{\phi }^{\mu }$, where $T$ is the stress-energy tensor. The poloidal components of these fluxes are the following \citep[e.g.,][]{Blandford1977MNRAS.179..433B}:
\begin{align}
\mathcal{E}^r&=-\frac{\omega  B_T}{4\pi\Sigma  \sin \theta} \psi_{,\theta },\quad
\mathcal{E}^{\theta }=\frac{\omega B_T}{4\pi\Sigma  \sin \theta } \psi_{,r},\\
\mathcal{E}^r&=\omega  \mathcal{L}^r,\quad
\mathcal{E}^{\theta }=\omega  \mathcal{L}^{\theta }.
\end{align}
So the amount of energy and angular momentum flux transported per unit magnetic flux is (counting both sides of the disk)
\begin{align}
\frac{d^2E}{dt d\psi}&=\omega(\psi) B_T(\psi),\\
\frac{d^2L}{dt d\psi}&=B_T(\psi).
\end{align}
Since we know the flux distribution on the disk, we can obtain the energy and angular momentum flux transported per unit radius on the disk surface \citep{Uzdensky2005ApJ...620..889U}
\begin{align}
\frac{d^2E}{dt dr}&=\omega(\psi) B_T(\psi)\frac{d\psi}{dr},\label{eq:EH}\\
\frac{d^2L}{dt dr}&=B_T(\psi)\frac{d\psi}{dr}.\label{eq:LH}
\end{align}
Noticing that at the black hole event horizon, using the Znajek boundary condition we can write 
\begin{equation}
\mathcal{E}^r=\frac{1}{4\pi}\omega \left(\Omega _H-\omega \right) \left(a^2+r_H^2\right)  \left(\frac{\psi_{,\theta }}{a^2  \cos ^2\theta +r_H^2}\right)^2,
\end{equation}
so $\mathcal{E}^{r}>0$ if $0<\omega<\Omega_H$, and $\mathcal{E}^{r}<0$ if $\omega>\Omega_H$. This means that if the black hole angular velocity is larger than that of the disk, the magnetic link is going to transport energy and angular momentum from the black hole to the disk; if the angular velocity of the black hole is less than that of the disk, energy and angular momentum goes from the disk to the black hole. When $a>0.3594$, $\Omega_H$ becomes larger than the disk angular velocity everywhere ($r\geq r_{\rm ISCO}$), so energy and angular momentum only go from the black hole to the disk. When $a<0.3594$, there is a radius $r_{\rm co}$ on the disk where the disk angular velocity equals to $\Omega_H$; the magnetic link connecting to the disk within $r_{\rm co}$ will transfer energy and angular momentum from the disk to the hole, while those outside $r_{\rm co}$ have the fluxes reversed \citep[e.g.,][]{Uzdensky2005ApJ...620..889U}.

\begin{figure}
  \centering
  	\includegraphics[width=\columnwidth]{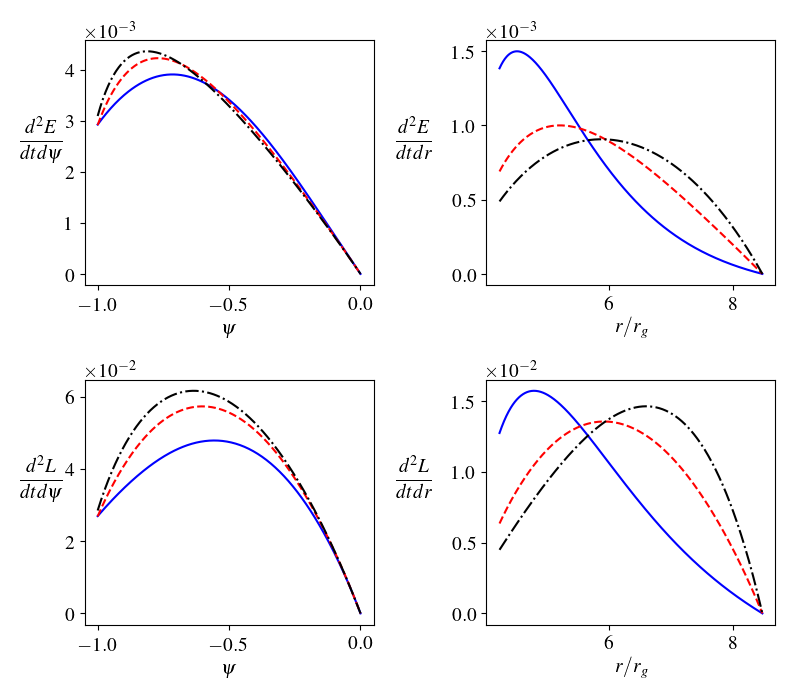}
  \caption{Energy and angular momentum flux deposited onto the disk by the linking magnetic flux loops, as a function of $\psi$ ($d^2E/dt d\psi$, left panels) and as a function of the disk radius $r$ ($d^2E/dt dr$, right panels), for three different cases corresponding to Figure \ref{fig:a0.5compare}. These all have a black hole spin $a=0.5$ and the last closed field line connects to the disk at $r_0=2r_{\rm ISCO}$. The difference is the flux distribution on the disk: blue solid line---monopole-like field at large distances; red dashed line---paraboloidal field; black dash-dotted line---vertical field. The magnetic flux has been normalized so that all three cases have the same amount of flux going through the black hole event horizon.}\label{fig:a0.5-Er-Lr}
\end{figure}

\begin{figure}
  \centering
  	\includegraphics[width=\columnwidth]{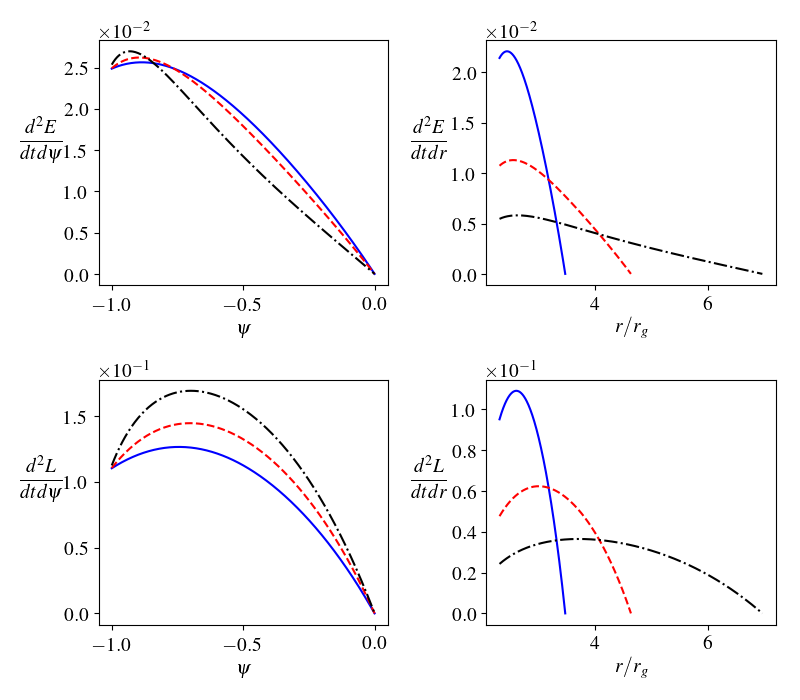}
  \caption{Energy and angular momentum flux deposited onto the disk by the linking magnetic flux loops, as a function of $\psi$ ($d^2E/dt d\psi$, left panels) and as a function of the disk radius $r$ ($d^2E/dt dr$, right panels), for three different cases where the black hole spin is $a=0.9$ and the magnetic flux is asymptotically paraboloidal, while the last closed field line connects to the disk at different radii: blue solid line---$r_0=1.5r_{\rm ISCO}$; red dashed line---$r_0=2r_{\rm ISCO}$; black dash-dotted line---$r_0=3r_{\rm ISCO}$. The magnetic flux has been normalized so that all three cases have the same amount of flux going through the black hole event horizon.}\label{fig:a0.9-Er-Lr}
\end{figure}

\begin{figure}
  \centering
  \includegraphics[width=\columnwidth]{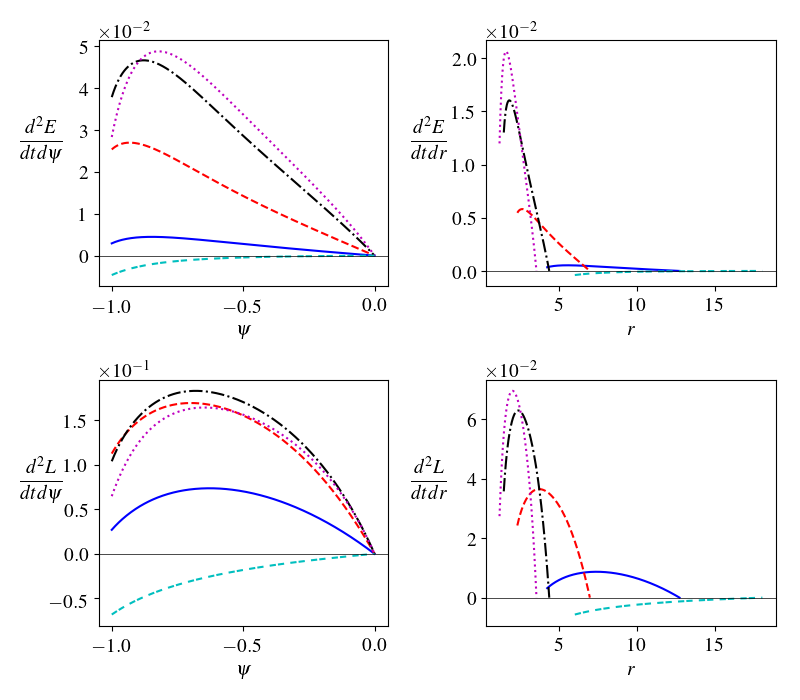}
  \caption{Energy and angular momentum flux deposited onto the disk by the linking magnetic flux loops, as a function of $\psi$ ($d^2E/dt d\psi$, left panels) and as a function of the disk radius $r$ ($d^2E/dt dr$, right panels), for five different cases similar to Figure \ref{fig:a-compare}. All five cases have asymptotically paraboloidal magnetic field, and the last closed field line connects to the disk at $r_0=3r_{\rm ISCO}$. The difference is the black hole spin: cyan double-dashed line---$a=0$; blue solid line---$a=0.5$; red dashed line---$a=0.9$; black dash-dotted line---$a=0.99$; magenta dotted line---$a=0.999$. The magnetic flux has been normalized so that all four cases have the same amount of flux going through the black hole event horizon.}\label{fig:rs3-Er-Lr}
\end{figure}

\begin{figure}
  \centering
  	\includegraphics[width=\columnwidth]{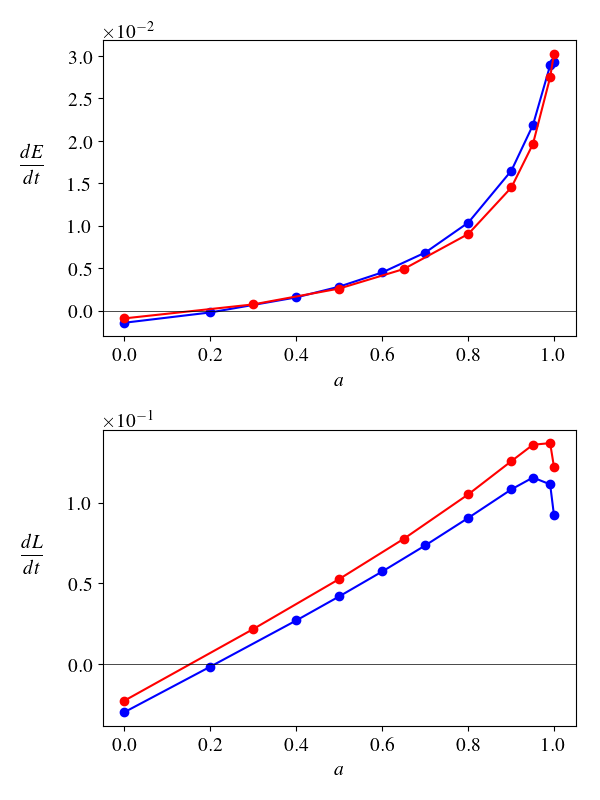}
  \caption{Total energy and angular momentum flux deposited onto the disk by the linking magnetic loops as a function of the black hole spin, for the cases where the magnetic field is asymptotically paraboloidal. Blue points have the last closed field line connecting to the disk at $r_0=2r_{\rm ISCO}$, while red points have $r_0=3r_{\rm ISCO}$. The magnetic flux has been normalized so that all the cases have the same amount of flux going through the black hole event horizon.}\label{fig:scaling}
\end{figure}

Figure \ref{fig:a0.5-Er-Lr} shows the comparison among three cases where both the black hole spin and the separatrix foot point are kept the same while only the flux distribution on the disk changes. We can see that when the closed zone is pushed down by the external pressure, the total amount of energy/angular momentum flux transported to the disk (integrated over $\psi$) slightly increases.

Figure \ref{fig:a0.9-Er-Lr} shows the comparison among three cases where both the black hole spin and the asymptotic magnetic field are kept the same while only the location of the separatrix foot point on the disk $r_0$ changes. It can be seen that although the total energy flux transported to the disk slightly decreases as $r_0$ increases, the total angular momentum flux increases.

Figure \ref{fig:rs3-Er-Lr} shows the comparison among five cases where both the asymptotic magnetic field and the location of the separatrix foot point on the disk in terms of $r_{\rm ISCO}$---$r_0/r_{\rm ISCO}$---are kept the same while only the black hole spin changes. We can see that the total energy transported to the disk increases with the black hole spin $a$; the total angular momentum flux first increases with $a$, then decreases at very large spins. To see this more clearly, we plot the total energy and angular momentum flux carried by the magnetic links (integrated over $\psi$) as a function of $a$ in Figure \ref{fig:scaling}. At relatively small spins, the angular momentum flux is almost linear in $a$, but for very large spins, it deviates from the trend and appears smaller.

As an order of magnitude estimation, suppose the magnetic flux threading the black hole event horizon is $\Phi$, then the available voltage is $V\sim\omega\Phi/2\pi\sim\epsilon_1\Phi a c/r_g$. Measuring $B$ in terms of the characteristic field $B_{\rm Edd}\equiv\sqrt{8\pi p_{\rm Edd}}=\sqrt{2L_{\rm Edd}/3cr_g^2}=\sqrt{m_pc^2/r_e^2r_g}=3.6\times10^5M_6^{-1/2}$ G \citep{phinney_theory_1983}, we get $V\sim1.6\times10^{19}\epsilon_1a (B/B_{\rm Edd})M_6\;{\rm V}$. So the power output from the black hole is $P\sim V^2/Z_0\sim\epsilon_2 \Phi^2a^2c^2/(r_g^2Z_0)\sim6.8\times10^{42}\epsilon_2a^2(B/B_{\rm Edd})^2M_6^2\;{\rm erg~s^{-1}}$, and the angular momentum flux is $\tau\sim P/\omega\sim\epsilon_3\Phi^2 a c/(r_gZ_0)$, where $r_g=GM/c^2$ is the gravitational radius, $Z_0=\mu_0 c=377\Omega$ is the impedance of the vacuum (in SI units), and from our numerical results $\epsilon_2\sim10^{-2}$, $\epsilon_3\sim10^{-1}$. 

Another point we would like to mention is that here we only consider Poynting flux. In reality, if the mass loading on the field line is significant, e.g., $\sigma<1$, then the total energy and/or angular momentum flux could be going in the opposite direction, namely, from the disk to the black hole, even if $0<\omega<\omega_H$ \citep[e.g.][]{Globus2013PhRvD..88h4046G}.

\subsection{Thermal emissivity profile of the disk---a simplistic approach}

If the energy flux transported by the magnetic link is dissipated at the disk into heat, it will change the thermal emissivity profile of the disk \citep[e.g.,][]{1999Sci...284..115V,Li2002ApJ...567..463L}. Let us first adopt the simplest treatment: the Poynting flux extracted from the black hole is fully deposited on the disk, without loss on the way; and we follow \citet{Li2002ApJ...567..463L}, by assuming that the disk is quasi-steady and thin, orbiting on nearly geodesic, circular orbits, and the angular momentum of the accreting matter, plus that injected by the magnetic link, is transported outward through viscous stress in the disk. Under these conditions, one can use the radial conservation laws to obtain the viscous stress and radiative flux as functions of radius $r$, while all the complexities involving physics of viscosity and radiation are left untouched in the vertical structure of the disk.

In the steady state, the mass accretion rate $\dot{M}$ is a constant. The radial equation for the angular momentum conservation can be written as
\begin{equation}
(-\dot{M} \ell+G)_{,r}+4 \pi r F \ell=L_H,
\end{equation}
where 
\begin{equation}
\ell=u_{\phi}=\frac{\frac{a^2 \sqrt{M}}{r^{3/2}}-\frac{2 a M}{r}+\sqrt{M r}}{\sqrt{\frac{2 a \sqrt{M}}{r^{3/2}}-\frac{3 M}{r}+1}}
\end{equation}
is the angular momentum of unit mass moving on direct, circular geodesic orbit; $G$ represents the viscous torque; $F$ is the energy flux radiated away from the surface of the disk, measured in the local rest frame of the disk material; $L_H$ is the angular momentum flux injected by the magnetic link: $L_H=d^2L/dt\, dr$ in Equation (\ref{eq:LH}). Here we neglect the energy and angular momentum carried away by the open field lines outside the closed zone. Similarly, the energy conservation equation is
\begin{equation}
(-\dot{M} \varepsilon+\omega G)_{,r}+4 \pi r F \varepsilon=E_H,
\end{equation}
where 
\begin{equation}
\varepsilon=-u_0=\frac{\frac{a \sqrt{M}}{r^{3/2}}-\frac{2 M}{r}+1}{\sqrt{\frac{2 a \sqrt{M}}{r^{3/2}}-\frac{3 M}{r}+1}}
\end{equation}
is the energy per unit mass moving on direct, circular geodesic orbit, $\omega=(a+r^{3/2}/M^{1/2})^{-1}$ is the angular velocity of the disk matter, and $E_H$ is the energy flux injected by the magnetic link, $E_H=d^2E/dt\, dr$ in Equation (\ref{eq:EH}). Noticing that $\varepsilon_{,r}-\omega \ell_{,r}=0$ and $E_H=\omega L_H$, we can combine the two equations and obtain
\begin{equation}
G=\frac{\varepsilon-\omega  \ell}{-\omega _{,r}}4\pi r F,
\end{equation}
and we are left with an ordinary differential equation for $G$ (or $F$). Here for boundary condition we assume $G(r_{\rm ISCO})=0$, namely, no viscous stress at the inner boundary of the disk. Then the differential  equation can be readily solved numerically, plugging in $L_H$ and $E_H$ from our Grad-Shafranov solutions.

\begin{figure}
	\centering
    \includegraphics[width=\columnwidth]{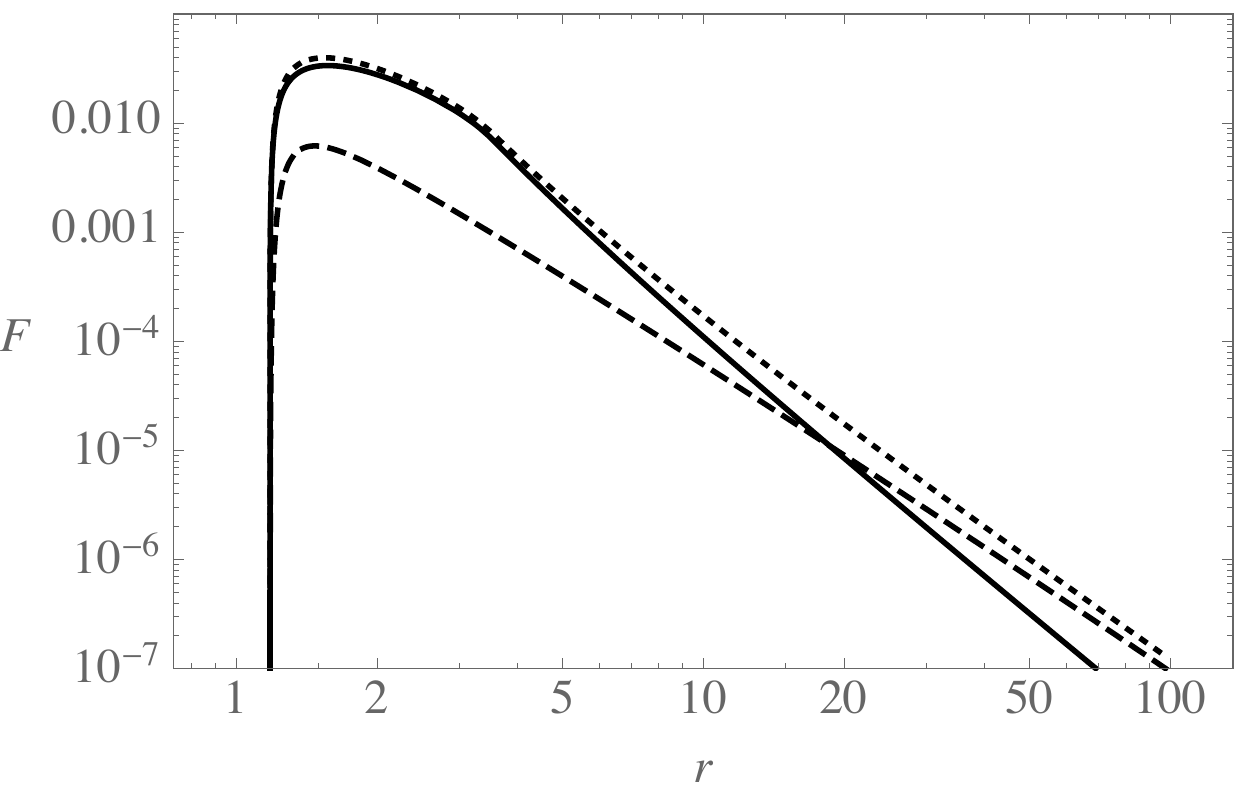}
    \caption{Thermal emissivity profile of a thin disk around a Kerr black hole with $a=0.999$. (1) Dashed line: standard disk as in \citet{Novikov1973blho.conf..343N}. (2) Solid line: magnetic link between the hole and the disk extends to $r_0=3r_{\rm ISCO}$, with $\dot{M}=0$ and arbitrary magnetic flux normalization. (3) Dotted line: similar to case (2) but with the same nonzero $\dot{M}$ as case (1).}
    \label{fig:a0.999-rs3-emissivity}
\end{figure}

\begin{figure}
	\centering        \hspace{0.6cm}\includegraphics[width=0.9\columnwidth]{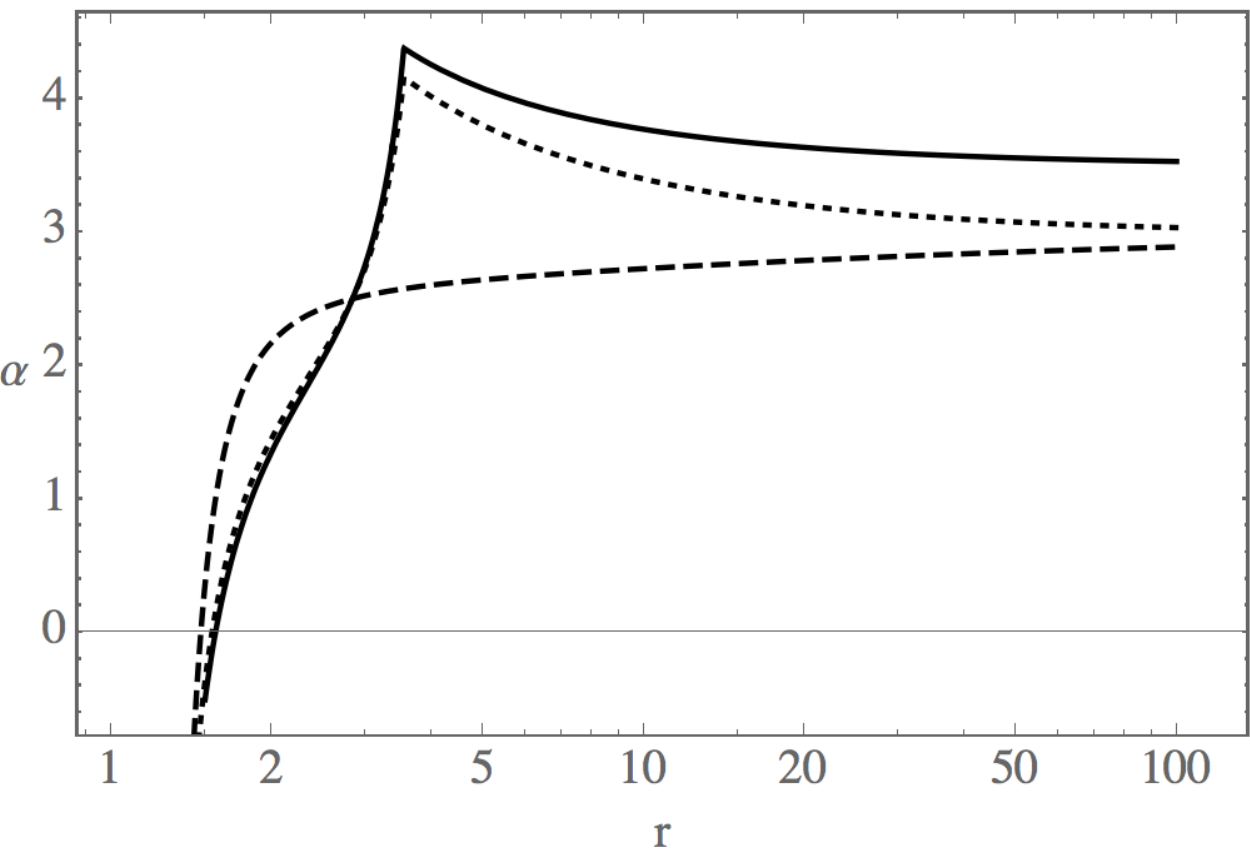}
    \caption{Emissivity index $\alpha\equiv -d\ln F/d\ln r$ for the three cases in Figure \ref{fig:a0.999-rs3-emissivity}.}
    \label{fig:a0.999-rs3-index}
\end{figure}

In Figure \ref{fig:a0.999-rs3-emissivity} we show the emissivity profile of one particular case, where the black hole spin is close to extremal: $a=0.999$, and compare it with the standard disk of \citet{Novikov1973blho.conf..343N}. When $\dot{M}=0$, we get nonzero radiation flux: in this case the energy source is the rotational energy of the black hole transported to the disk through the magnetic link, and at large disk radii, the emissivity index $\alpha\equiv -d\ln F/d\ln r$ approaches 3.5, consistent with the result of \citet{Li2002ApJ...567..463L}. As a comparison, a standard disk of \citet{Novikov1973blho.conf..343N} can only have nonzero radiation when $\dot{M}\neq0$, and the emissivity index approaches 3 at large radii. When the magnetic link coexists with accretion, it turns out that both the viscous torque $G$ and the radiation flux $F$ are enhanced in the region where the disk-hole link is depositing energy and angular momentum, and this enhancement extends to larger radii beyond the separatrix foot point $r_0$. The emissivity index $\alpha$ approaches the asymptotic value of 3 similar to a standard disk at large distances from the black hole (Figure \ref{fig:a0.999-rs3-index}). 

Note that the thermal emissivity profile we calculated above is different from the reflection fluorescent line emissivity profile discussed by \citet{Wilkins2011MNRAS.414.1269W,Wilkins2015MNRAS.449..129W}. 
It is likely that if the Poynting flux extracted from the black hole has significant dissipation on the axis before it is transported to the disk, we will have a compact emitting source at the dissipation site that irradiates the disk, thus producing a reflection emissivity profile that resembles the models in \citet{Wilkins2011MNRAS.414.1269W}.


\section{Discussion}\label{sec:discussion}
\subsection{Stability of the solutions}
We have shown that a steady state solution with disk-hole linking field lines can be obtained using a relaxation method in certain parameter regimes, and there exists a maximum extent of the closed zone, depending on the black hole spin and the flux distribution on the disk. A question that may arise is whether the obtained solutions are stable. 

We should consider three approaches. In the first, we suppose that the solution is constrained to remain axisymmetric and only axisymmetric perturbations are considered.
\citet{2002ApJ...565.1191U} showed that for a given boundary condition, the magnetostatic, non-relativistic force-free equilibrium equation $(\nabla\times\pmb{B})\times\pmb{B}=0$ can have two branches of solutions: one stable and the other unstable. In that case, the characteristic speed for information propagation is necessarily infinite, so when the actual dynamics is taken into account, some of these solutions turn out to be acausal and dynamically unstable. In our case, the finite characteristic speeds are already included, whose direct consequence is the existence of the light surfaces. In the force-free limit, the Alfv\'{e}n critical surfaces coincide with the light surfaces and the fast critical surfaces coincide with the event horizon and the infinity\footnote{When one considers an actual MHD system with plasma inertia properly included, the Alfv\'{e}n critical surfaces lie in between the inner and outer light surfaces; the slow magnetosonic surfaces lie close to the launching point/stagnation point; the fast magnetosonic surface lies between the inner light surface and the event horizon, or between the outer light surface and infinity \citep{phinney_theory_1983,Takahashi1990}.}. Our solutions have the closed zone within the sub-Alfv\'{e}nic region (except for the small portion close to the event horizon), so they are physically causal. But the stability may still rely on a time dependent simulation to tell.

In the second approach, we relax the constraint of axisymmetry and allow non-axisymmetric signals to communicate inward and outward. The axisymmetric light surfaces are insufficient to understand the whole problem and stability of a given equilibrium can only be demonstrated through a thorough examination of the fate of all perturbations to an initial value problem. Mathematically, this task is daunting. The pragmatic approach is through non-axisymmetric, time-dependent simulations that start from these equilibria as we shall discuss in a future paper. 

The third approach is to also include plasma effects. Initially this can be done in the MHD limit when it is the inertia of the fluid that is of most importance. However, ultimately this is a kinetic problem where highly non-thermal particle distribution functions and coupling of radiation become relevant as dissipative agents.

What we envisage is that, most of the solutions may be stable under axisymmetry constraint, but they may be unstable in 3D, as we discuss below.

\subsection{3D reconnection at the null point}\label{subsec:3D reconnection}

In the above configurations we considered, the point where the last closed field line encounters the pole is a null point of the magnetic field ($B=0$ here). We have seen that for typical black hole spins and typical flux distribution on the disk, the null point is located at a few $r_g\equiv GM/c^2$ above the black hole. Magnetic reconnection can happen at this point, and it is fully 3D, different from the typical 2D X-type reconnection well studied in the literature. In particular, it bears some resemblance to the null point reconnection being investigated in the solar coronal jet context \citep[e.g.,][]{2018ApJ...852...98W}.


Neglect the effect of accretion at the moment. Imagine there is a non-axisymmetric perturbation such that the closed zone tilts slightly away from the axis, and the null point of the closed zone is displaced from the null point of the open field zone. The null point of the closed zone will then encounter a higher stress region of the external open field and gets pushed inward, while the section of the closed zone that touches the null point of the open zone will push outward. In a word, the stress imbalance facilitates further tilting of the axis of the closed zone, like the tilting of a spheromak in a uniform external field \citep{Rosenbluth1979NucFu..19..489R, bellan_spheromaks_2000}. But a few processes may keep the tilting instability at some finite amplitude, without running away. Firstly, The open field lines that push down on the null point may encounter the opposite closed field lines from the other side and reconnect with them. This leads to an increase of the closed field on one side and a decrease on the other side, and vice versa for open flux. At the same time, the spin of the black hole will try to straighten up the axis; it may eventually deposit too much stress in the region where an excess of closed field lines exists, leading to some of these field lines open up in a manner like coronal mass ejection. The loose end of the opened up field line on the black hole may eventually reconnect with an opposite open field from the disk to get back to a closed loop. So the closed zone can still maintain its integrity. These processes may shuffle around the open flux on the disk as well as the closed flux but may not be able to change the total amount of closed flux. The dissipated magnetic energy is eventually compensated by the rotational energy of the black hole (and possibly the kinetic energy of the accreted matter, if the mass loading on the field lines is significant). The details of the time-dependent evolution will be studied in a forthcoming publication.

We can make some simple estimation of the energetics based on this scenario. Suppose the region surrounding the separatrix is continuously subject to reconnection, and the reconnection speed is $v_{\rm rec}\sim 0.1c$. Then the power dissipated at the separatrix is $P_{\rm diss}\sim B^2 r^2v_{\rm rec}/8\pi$, where $r$ is the radial location of the separatrix point, typically a few $r_g$. Measuring $B$ in terms of the characteristic field $B_{\rm Edd}$, we get
\begin{equation}\label{eq:P_dissipation}
\frac{P_{\rm diss}}{L_{\rm Edd}}\sim\frac{1}{12\pi}\left(\frac{v_{\rm rec}}{c}\right)\left(\frac{r}{r_g}\right)^2\left(\frac{B}{B_{\rm Edd}}\right)^2.
\end{equation}
To compare with the disk emission, suppose the accretion rate is $\dot{M}$ and the disk luminosity is on the order of $L_{\rm disk}=\epsilon_r \dot{M}c^2$, where $\epsilon_r\sim0.1$ is the radiation efficiency. Suppose that the maximum magnetic field is achieved such that near the event horizon, the magnetic pressure equals to the ram pressure of the accreted matter, namely $B^2\sim \dot{M}c/r_g^2$, then we get
\begin{equation}
\frac{P_{\rm diss}}{L_{\rm disk}}\sim\frac{1}{8\pi\epsilon_r}\left(\frac{v_{\rm rec}}{c}\right),
\end{equation}
so the power dissipated due to reconnection near the separatrix could in principle get close to the disk thermal luminosity.


\subsection{Flux transport on the disk and formation of closed field region}
Imagine that the angular momentum in the disk (including the amount injected by the black hole through the linked magnetic field) can be efficiently transported outward by small scale turbulence in the disk, so that the disk material gets efficiently accreted. Then the closed zone as shown above can only exist for a time scale $\sim$ size of the loop divided by the accretion speed. If the open field line region outside the closed loops is roughly unidirectional, then when the foot points of the closed loops are swallowed by the black hole, we are left with unidirectional field on the black hole and the disk. However, the disk may actually be more turbulent and carry with it different signs of fluxes with variable coherent length scales. This may be a natural result of magneto-rotational instability (MRI) in the disk \citep[e.g.,][]{2010ApJ...713...52D}. In this case, when flux of opposite sign is advected toward the black hole, it will reconnect with that on the hole, and form a closed disk-hole link \citep{Parfrey2015MNRAS.446L..61P}. The reconnection process involved is violent, releasing most of the magnetic energy contained in the open magnetic field on the hole---it may be regarded as a big flare. When the closed loops form, they start out with the largest extent, then as the disk foot points get gradually accreted, the extent of the loop decreases, and the null point where 3D reconnection happens also gets closer to the black hole. During the slow shrinking of the closed zone, the dissipation process as discussed in \S\ref{subsec:3D reconnection} continues to operate all the time and heats up the plasma around the null point. This whole cycle happens again and again as different signs of fluxes are advected to the black hole by the disk, and we may expect a highly variable corona region above the black hole most of the time, depending on the coherent length scale of the flux distribution.

Another way of replenishing the disk-hole linking magnetic field is that the disk may launch a mass-loaded wind along the open flux tubes; due to the hoop stress of the toroidal magnetic field, the wind will be collimated along the axis. Now either because the material launched from the field lines near the axis might not reach the escape velocity (the effective potential maximum is located at relatively large distances near the axis), or because the hoop stress squeezes strongly at some point forcing matter to move both upward and downward, the material loaded on the near-axis field lines may turn around and fall into the black hole, dragging the field lines with it. This matter can pin some magnetic loops onto the black hole, and as it falls through the magnetic field, it causes dissipation and heating near the separatrix too.

\subsection{On the force-free approximation and possible effects of mass loading}
In this paper, we mostly considered the magnetic configurations in the force-free limit. We can estimate the realistic magnetization values from observational data.

Take the narrow line Seyfert I galaxy NGC 4151 as an example. Here the black hole mass is $M=(4.5\pm0.5)\times 10^{7}M_{\odot}$ \citep{2006ApJ...651..775B}, so $r_g=6.7\times 10^{12} \rm{cm}\approx2.2\times10^2$ light seconds, and $L_{\rm Edd}=5.7\times 10^{45}\,\rm{erg}\,\rm{s}^{-1}$. The X-ray luminosity $L_X\sim 5\times 10^{43}\,\rm{erg}\,\rm{s}^{-1}\sim0.01L_{\rm Edd}$ \citep{1996MNRAS.283..193Z}. The corona is located at $r\sim6r_g$ based on Fe line reverberation mapping\citep{2012MNRAS.422..129Z}.
From Equation (\ref{eq:P_dissipation}) we need a magnetic field of $B\sim 0.3 B_{\rm Edd}\sim 5.4\times10^4$ G.
If most of the hard X-ray emission is produced by thermal Comptonization, we need the Compton y parameter to be larger than 1, namely
\begin{equation}
    y_{\rm NR}=\frac{4kT}{m_ec^2}\rm{Max}(\tau_{\rm es},\tau_{\rm es}^2)>1,
\end{equation}
where $\tau_{\rm es}=n\sigma_T r$ is the Thomson scattering optical depth of a photon traversing the plasma, $n$ is the electron density, and $T$ is the electron temperature. Since $\tau_{\rm es}\gtrsim1$, we have 
\begin{equation}
    n\sim 4\times10^{10}y_{\rm NR}^{1/2}\left(\frac{T}{100\,{\rm keV}}\right)^{-1/2}\left(\frac{r}{6r_g}\right)^{-1}.
\end{equation}
With these parameters, the electron magnetization is
\begin{equation}
    \sigma_e\equiv\frac{B^2}{4\pi n m_e c^2}\sim 7\times 10^3y_{\rm NR}^{-1/2}\left(\frac{T}{100\,{\rm keV}}\right)^{1/2}\left(\frac{r}{6r_g}\right).
\end{equation}
If there are equal numbers of electrons and protons, the magnetization of the plasma would be reduced to order unity. However, it is more likely that there is only a small fraction of protons while the plasma is pair dominated \citep{1996MNRAS.283..193Z}, so it may well be highly magnetized in the corona region. The force-free limit should be a good approximation in such a situation.

Field lines satisfying the conditions described in \S \ref{sec:mass loading} could have mass flow along them. It is expected that small amount of mass loading will not change the force-free magnetic configuration qualitatively, while large amount of mass loading could modify the causal conditions and significantly alter the field structure.  One can imagine a situation where the plasma dynamics in the corona is largely governed by the magnetic field, which may be produced by small scale currents in the accretion disk (likely a result of MRI and dynamo processes in the disk) thus varies on relatively small length scales. 
The local magnetic field at the disk surface changes in a fraction of an orbital period. Closed field loops open up and vice versa \citep[e.g.,][]{UzdenskyGoodman2008ApJ...682..608U,Parfrey2015MNRAS.446L..61P}. The mass loading on some flux tubes might be large while neighboring flux magnetic field might be force-free. Causal connection might not happen directly along a closed flux tube but could still happen indirectly through the environment. A high resolution, time-dependent MHD code would be needed to explore this case.

\section{Conclusions}\label{sec:conclusion}

In this paper we have obtained simple axisymmetric models of a general relativistic, force-free field near a Kerr black hole, where there are flux tubes linking the hole and the disk (closed zone), confined by external open field from the disk. We find that the extent of the closed zone depends on the black hole spin as well as the pressure from the open field zone. Increasing the relative strength of the open field helps to confine the closed flux tubes, while increasing black hole spin typically induces stronger toroidal field on the closed flux tubes that pushes against the external confinement \citep[consistent with][]{Uzdensky2005ApJ...620..889U}. For a typical disk field flux distribution, the maximal extent of the closed zone decreases with the black hole spin, and can be about a few $r_g$ for high spins.

The disk-hole linking field lines can transport energy and angular momentum from the black hole to the disk or vice versa, depending on the relative angular velocity. For high spins, energy is extracted from the black hole. Assuming all this energy is deposited onto the disk and dissipated into heat there, we find that the thermal emissivity profile of the disk shows more concentration at the inner region than the standard disk.

The separatrix layer between the closed field zone and open field zone could be a potential site for dissipation, due to their interaction/competition in pressure balance. The stability and possible dissipation will be studied in a forthcoming publication using time dependent force-free simulations.

\section*{Acknowledgements}

We thank Alex Chen, Erin Kara, Kyle Parfrey, Anatoly Spitkovsky, and especially, Dmitri Uzdensky for helpful discussion. We also thank the anonymous referee for constructive comments on the manuscript.
YY acknowledges support from the Lyman Spitzer, Jr. Postdoctoral Fellowship awarded by the Department of Astrophysical Sciences at Princeton University. RDB ackowledges support by the Miller Institute and the Simons Foundation. DRW is supported by NASA through Einstein Postdoctoral Fellowship grant number PF6-170160, awarded by the \textit{Chandra} X-ray Center, operated by the Smithsonian Astrophysical Observatory for NASA under contract NAS8-03060. 




\bibliographystyle{mnras}
\bibliography{ref}



\appendix

\section{Numerical method for solving the force-free Grad-Shafranov equation}\label{sec:method}

Throughout the paper we use Boyer-Lindquist coordinates
\begin{align}
ds^2&=-\left(1-\frac{2 M r}{\Sigma }\right)dt^2 +\frac{\Sigma }{\Delta }dr^2+\Sigma d\theta^2+\frac{\mathcal{A}  \sin ^2\theta }{\Sigma }d\phi^2 \nonumber\\
&-\frac{ 4 a M r \sin ^2 \theta }{\Sigma }dt\,d\phi,
\end{align}
where $\Sigma =r^2+a^2  \cos ^2\theta$, $\Delta =r^2-2 M r+a^2$, $\mathcal{A}=\left(r^2+a^2\right)^2-\Delta a^2 \sin ^2 \theta$.

For an axisymmetric, steady state, force-free configuration, the flux function $\psi\equiv A_{\phi}$ (which is the $\phi$ component of the 4-vector potential $A_{\mu}$) satisfies the so-called Grad-Shafranov equation \citep[e.g.,][]{Blandford1977MNRAS.179..433B}
\begin{align}\label{eq:GS}
\frac{B_T(\psi) B_T'(\psi)}{\Delta\sin ^2\theta}&=\frac{1}{\sqrt{-g}}\left[\left(\frac{\sqrt{-g}g^{rr} K}{\Delta \sin ^2 \theta}\psi _{,r}\right)_{,r}+\left(\frac{\sqrt{-g}g^{\theta \theta } K}{\Delta \sin ^2 \theta}\psi _{,\theta }\right)_{,\theta }\right]\nonumber\\
&-\frac{(\nabla \psi )^2 \omega'(\psi) \left(g_{0 \phi }+\omega(\psi)  g_{\phi \phi }\right)}{\Delta  \sin ^2 \theta },
\end{align}
where $K=g_{tt}+2  g_{t\phi }\omega(\psi) +g_{\phi \phi }\omega ^2(\psi) $ and $(\nabla \psi )^2=g^{\theta \theta }\left(\partial \psi /\partial \theta \right)^2+g^{rr}\left(\partial \psi /\partial r\right)^2$. The toroidal magnetic field $B_T=\sqrt{-g} g^{r r} g^{\theta \theta } (A_{\theta ,r}-A_{r,\theta }) $ (or equivalently the poloidal current $I=B_T$) and field line angular velocity $\omega$ are functions of $\psi$. A solution to this equation should determine $\psi(r,\theta)$ and the functional forms $B_T(\psi)$, $\omega(\psi)$ at the same time.

In terms of $\psi(r,\theta)$, (\ref{eq:GS}) is an elliptic partial differential equation. Note that the equation is singular on the surface(s) where $K=0$---the light surface(s). Although the equation does not change its character when going across the light surface, the smoothness condition at the light surface imposes constraint on the flux functions. In order for the solution to pass smoothly through the light surface, we need to satisfy the following condition, which is obtained by taking the limit $K\to 0$ in Equation (\ref{eq:GS}):
\begin{align}\label{eq:light-surface-I}
0&=B_TB_T'-\left[g^{rr}\psi_{,r}K_{,r}+g^{\theta\theta}\psi_{,\theta}K_{,\theta}-(\nabla\psi)^2\omega'(g_{t\phi}+\omega g_{\phi\phi})\right]\nonumber\\
&\equiv Q,
\end{align}
and on the light surface $\mathcal{S}$, the second derivative of $\psi$ is determined from
\begin{equation}\label{eq:light-surface-derivative}
g^{rr}\frac{\partial ^2\psi }{\partial r^2}+g^{\theta \theta }\frac{\partial ^2\psi }{\partial \theta ^2}=\frac{\left.\frac{\partial Q}{\partial r}\right|_{\mathcal{S}}}{\left.\frac{\partial K}{\partial r}\right|_{\mathcal{S}}}.
\end{equation}
The black hole event horizon and the infinity are also singular surfaces/points of Equation ($\ref{eq:GS}$), but it turns out that the solution only needs to satisfy certain regularity conditions at these locations \citep[e.g.,][]{Komissarov2004MNRAS.350..427K, Uzdensky2005ApJ...620..889U, Nathanail2014ApJ...788..186N}.

For our configurations, the field line angular velocity $\omega$ is given by the Keplerian angular velocity of the foot point on the disk, and $B_T(\psi)$ can be obtained from the smoothness condition (\ref{eq:light-surface-I}) at the light surface---each field line passes through one and only one light surface, which is just sufficient and necessary to constrain $B_T(\psi)$. One more condition is needed for a well defined problem: that is the boundary condition for $\psi$. Here the boundary conditions at $\theta=0$ and $\theta=\pi/2$ are important; boundary conditions at the event horizon $r=r_H$ and infinity $r=\infty$ coincide with the regularity condition and turn out to be unimportant for the final solution.

Systems like (\ref{eq:GS}) have been studied by a few authors before \citep[e.g.][]{Contopoulos1999ApJ...511..351C, Uzdensky2005ApJ...620..889U, Timokhin2006MNRAS.368.1055T, Nathanail2014ApJ...788..186N}. Here we describe a simple but robust relaxation scheme we developed ourselves based on these previous work.

To proceed, we write Equation (\ref{eq:GS}) into the standard form of second order partial differential equations:
\begin{align}
K\left(g^{rr}\frac{\partial^2 \psi}{\partial r^2}+g^{\theta\theta}\frac{\partial^2 \psi}{\partial \theta^2}\right)+C_r(r,\theta,\omega)\psi_{,r}+C_{\theta}(r,\theta,\omega)\psi_{,\theta}\nonumber\\
=B_TB_T'-(\nabla\psi)^2\omega'(g_{t\phi}+\omega g_{\phi\phi}),
\end{align}
or $\mathcal{L}\psi=q$, where $\mathcal{L}$ denotes the differential operator on the left hand side and $q$ denotes the source term on the right hand side.
The basic idea of solving this elliptic equation is to consider the following diffusion equation instead:
\begin{equation}\label{eq:diffusion}
\frac{\partial\psi}{\partial t}=\mathcal{L}\psi-q.
\end{equation}

In practice we make the change of variable from $r$ to $x=r/(1+r)$ so the infinity is brought back to $x=1$. We discretize the equation on a uniformly spaced 2D $x-\theta$ grid, start with an initial trial solution, and use the successive overrelaxation method \citep{press_numerical_1999} to let the solution evolve toward the final steady one. Since this is a nonlinear differential equation, the coefficient terms need to be updated every time step. For force-free field at large radii, sometimes a better non-linear solver is needed to ensure stability, for example, adding a midpoint in the time step could help a lot. Also due to the presence of the light surfaces, some additional care needs to be taken. Firstly, at the light surface, Equation (\ref{eq:GS}) breaks down, we use Equation (\ref{eq:light-surface-derivative}) instead. The way we do this is to first locate the light surface by linear interpolation on the grid (we find the radial location of the light surface for each $\theta$), then replace the coefficient terms on the adjacent grid points with that of Equation (\ref{eq:light-surface-derivative}). Secondly, at every time step, we use the smoothness condition (\ref{eq:light-surface-I}) at the light surface to determine $B_TB_T'$ as a function of $\psi$. This is also done for each $\theta$, which gives us a discrete list of $B_TB_T'$ value for each corresponding $\psi$. We then use a high order polynomial to fit $B_TB_T'$ as a function of $\psi$, and use this obtained functional form over the entire region in the next time step to calculate the coefficient terms. With these measures, the relaxation procedure can successfully converge to the steady state solution, if it does exist.


\bsp	
\label{lastpage}
\end{document}